\documentclass[pdflatex,sn-basic]{sn-jnl}
\usepackage{graphicx}%
\usepackage{multirow}%
\usepackage{amsmath,amssymb,amsfonts}%
\usepackage{amsthm}%
\usepackage{mathrsfs}%
\usepackage[title]{appendix}%
\usepackage{xcolor}%
\usepackage{textcomp}%
\usepackage{manyfoot}%
\usepackage{booktabs}%
\usepackage{algorithm}%
\usepackage{algorithmicx}%
\usepackage{algpseudocode}%
\usepackage{listings}%
\usepackage{float}
\usepackage{amsmath}
\usepackage{adjustbox}
\usepackage{threeparttable}
\usepackage{longtable}
\usepackage{amsfonts}
\usepackage{url}
\usepackage{lscape}     
\usepackage{booktabs}      
\usepackage{tabularx}      
\usepackage{ragged2e}      
\newcolumntype{Y}{>{\RaggedRight\arraybackslash}X}
\newcolumntype{Z}{>{\RaggedRight\arraybackslash}p{3cm}} 

\theoremstyle{thmstyleone}%
\theoremstyle{thmstyletwo}%

\theoremstyle{thmstylethree}%

\begin{document}

\title{A Computational Perspective on NeuroAI and Synthetic Biological Intelligence}


\author[1,2,3]{\fnm{Dhruvik} \sur{Patel}}\email{pateld12@rpi.edu}
\equalcont{These authors contributed equally to this work.}

\author[1,2]{\fnm{Md Sayed} \sur{Tanveer}}\email{azamm@rpi.edu}
\equalcont{These authors contributed equally to this work.}

\author[4]{\fnm{Jesus} \sur{Gonzalez-Ferrer}}\email{jesusgzlezferrer@gmail.com}

\author[5]{\fnm{Alon} \sur{Loeffler}}\email{alon@corticallabs.com}

\author*[5,6]{\fnm{Brett J.} \sur{Kagan}}\email{brett@corticallabs.com}

\author*[4]{\fnm{Mohammed A.} \sur{Mostajo-Radji}}\email{mmostajo@ucsc.edu}

\author*[1,2]{\fnm{Ge} \sur{Wang}}\email{wangg6@rpi.edu}

\affil*[1]{\orgdiv{Department of Biomedical Engineering}, \orgname{Rensselaer Polytechnic Institute}, \orgaddress{\city{Troy}, \postcode{12180}, \state{NY}, \country{USA}}}

\affil[2]{\orgdiv{Center for Biotechnology and Interdisciplinary Studies}, \orgname{Rensselaer Polytechnic Institute}, \orgaddress{\city{Troy}, \postcode{12180}, \state{NY}, \country{USA}}}

\affil[3]{\orgname{Albany Medical College}, \orgaddress{\city{Albany}, \postcode{12208}, \state{NY}, \country{USA}}}

\affil[4]{\orgdiv{Genomics Institute}, \orgname{University of California Santa Cruz}, \orgaddress{\city{Santa Cruz}, \postcode{95064}, \state{CA}, \country{USA}}}

\affil[5]{\orgname{Cortical Labs}, \orgaddress{\city{Melbourne}, \state{VIC}, \country{Australia}}}

\affil[6]{\orgdiv{Department of Biochemistry and Pharmacology}, \orgname{The University of Melbourne}, \orgaddress{\city{Melbourne}, \state{VIC}, \country{Australia}}}

\abstract{NeuroAI is an emerging field at the intersection of neuroscience and artificial intelligence, where insights from brain function guide the design of intelligent systems. A central area within this field is synthetic biological intelligence (SBI), which combines the adaptive learning properties of biological neural networks with engineered hardware and software. SBI systems provide a platform for modeling neural computation, developing biohybrid architectures, and enabling new forms of embodied intelligence. In this review, we organize the NeuroAI landscape into three interacting domains: hardware, software, and wetware. We outline computational frameworks that integrate biological and non-biological systems and highlight recent advances in organoid intelligence, neuromorphic computing, and neuro-symbolic learning. These developments collectively point toward a new class of systems that compute through interactions between living neural tissue and digital algorithms.}




\maketitle

\section{Introduction}

The emergence of NeuroAI, particularly synthetic biological intelligence (SBI), represents a paradigm shift at the intersection of neuroscience, artificial intelligence, engineering, and biotechnology. This field leverages the inherent computational capabilities of biological neural systems for engineered intelligent behaviors. A key approach involves the development of \textit{in vitro} neural cultures and neural organoids, which recapture structural and functional features of the developing brain and enable the study of biological learning, memory, and reasoning processes \cite{kagan2023technology, smirnova2023organoid}. This multidisciplinary convergence enhances the potential of each contributing field while introducing new theoretical, technological, ethical, and regulatory challenges.

The pursuit of NeuroAI and SBI reflects an effort to understand and harness the computational power of biological neural networks (BNNs). By integrating living neural cultures with electronic interfaces, researchers emulate and potentially extend the brain's information-processing capabilities. This integration opens new opportunities for biomedical research, personalized medicine, and hybrid computing systems \cite{li2020organoid, tejavibulya2016personalized}. These platforms may support disease modeling, therapeutic screening, and the investigation of neural mechanisms underlying cognition, decision-making, and action \cite{haring2017microphysiological, wang2018modeling}.

The milestones enabling the development of SBI span more than a century, with contributions from both biological and computational sciences. Early biological advances include the first \textit{in vitro} tissue cultures in the late 19th century \cite{PHAM2018449} and the demonstration that neurons can be cultured \textit{ex vivo} in the early 20th century \cite{harrison1910outgrowth}. The introduction of multielectrode arrays (MEAs) in the 1970s provided a means to record neural activity and study circuit dynamics \cite{pine2006history}. The isolation of embryonic stem cells (ESCs) \cite{thomson1998embryonic} and the development of induced pluripotent stem cells (iPSCs) \cite{takahashi2006induction} enabled scalable, patient-specific generation of neural tissue. Recognizing the various limitations of two-dimensional cultures, researchers developed three-dimensional brain organoids in mouse and human models, leading to the onset of organoid intelligence (OI) \cite{eiraku2008self, kadoshima2013self, lancaster2013cerebral, lancaster2014generation}. These 3D constructs can faithfully replicate various aspects of \textit{in vivo} neural development, including cellular diversity and spatial organization, which are essential for the formation of functional neural networks \cite{li2020organoid, tejavibulya2016personalized}. 

In parallel, computational science has undergone transformative progress. Theoretical foundations were laid by early work on probabilistic models such as Markov chains \cite{chung1967markov, 2004TheLA, krogh1994hidden}, neural representations \cite{mcculloch1943logical}, the Turing test \cite{Turing1950computing}, and the 1956 Dartmouth Conference that launched the field of artificial intelligence \cite{frana2021encyclopedia}. Reinforcement learning approaches such as Temporal Difference Learning \cite{Sutton1988LearningTP} and Q-learning \cite{Watkins1992Qlearning} established formal methods for decision-making. The 21st century witnessed the emergence of deep learning, including convolutional networks like AlexNet \cite{NIPS2012_c399862d} and transformer-based architectures \cite{NIPS2017_3f5ee243}, culminating in today’s large language models \cite{NEURIPS2020_1457c0d6}. At the same time, neuroscientific theories such as Global Workspace Theory \cite{baars2005global}, Integrated Information Theory \cite{tononi2012integrated}, Predictive Coding \cite{kilner2007predictive}, and Active Inference \cite{friston2016active} have inspired biologically grounded strategies for developing more flexible and interpretable AI systems. This convergence is exemplified by recent NeuroAI frameworks that draw from both neuroscience and computer science, including neuro-symbolic systems that combine rule-based logic with neural learning \cite{garcez2023neurosymbolic, sheth2023neurosymbolic}, and neuromorphic hardware that emulates the spiking, event-driven architecture of biological neurons \cite{yang2020neuromorphic, ivanov2022neuromorphic, abdallah2022neuromorphic}.

Here, we examine the emerging field of SBI and NeuroAI with a focus on its computational foundations. We review key technological and conceptual advances, algorithms, and architectures that connect biological and artificial forms of intelligence. By highlighting these integrative efforts, we provide a comprehensive framework for understanding the current state and future directions of this rapidly evolving field.

\section{Biological Neural Networks (BNNs): The Foundation of Bio-intelligence and NeuroAI}

A central aim of NeuroAI is to design artificial or bio-engineered systems that mimic the structure and function of brains. This requires a clear understanding of the key characteristics of biological neural networks (BNNs). With the emergence of numerous recent breakthroughs in automated "smart" systems in the field of artificial intelligence (AI), which is based on artificial neural networks (ANNs), it is essential that we also discuss how BNNs differ from ANNs.

\label{sec:bnn_characteristics}

\subsection{Complexity of a Single Biological Neuron}

ANNs are often described as “brain-inspired,” with perceptrons serving as analogous to biological neurons \cite{block1962perceptron}. Yet, a real neuron performs significantly more complex computations than the simple weighted sum and threshold used in a perceptron.

\begin{figure}
    \centering
    \includegraphics[width=1\linewidth]{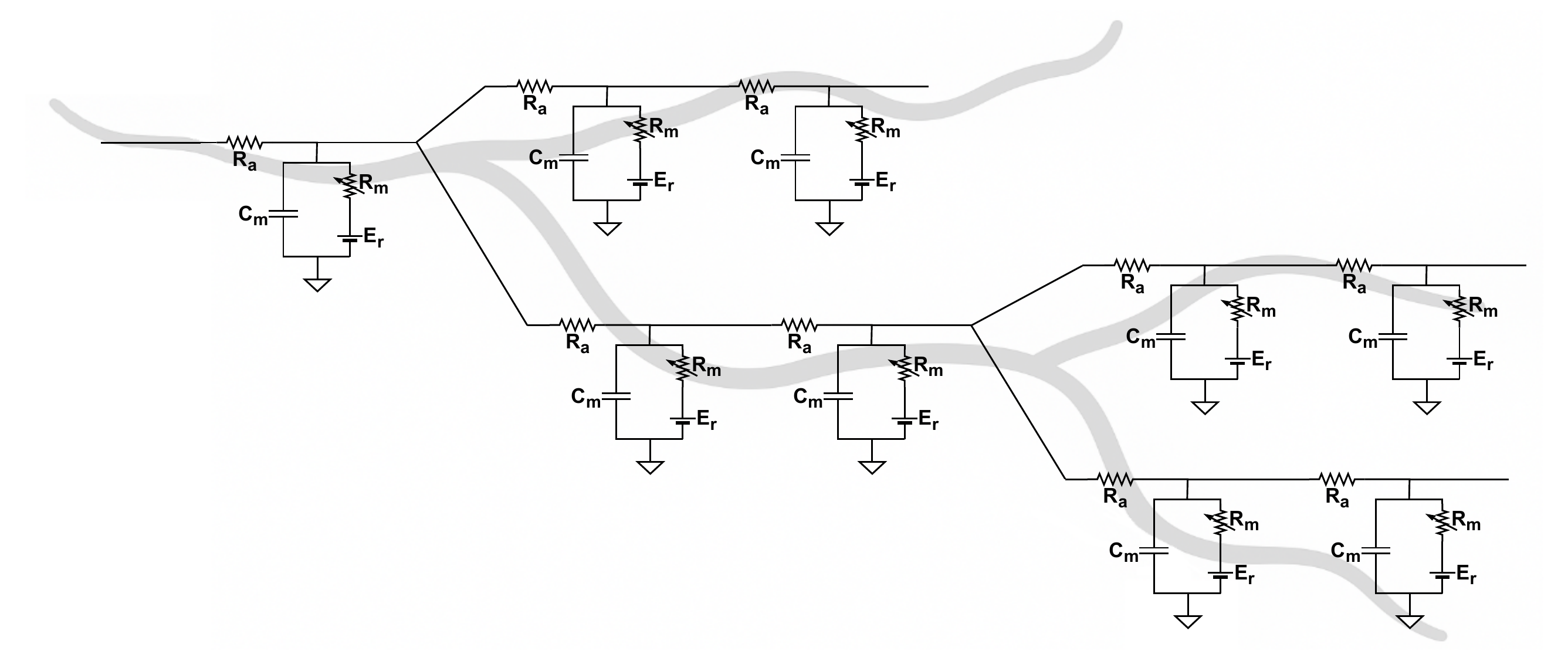}
    \caption{Representation of dendrite branches as electrical circuits with active elements.}
    \label{fig:dendrite_circuit}
\end{figure}

First, dendritic trees act not as single summing points but as distributed filter banks, as shown in Figure \ref{fig:dendrite_circuit}. Each branch approximates a leaky RC cable with distinct membrane resistance ($R_m$), axial resistance ($R_a$), and capacitance ($C_m$). Synaptic inputs undergo location- and time-dependent attenuation, with distal signals arriving more weakly and slowly than proximal ones \cite{london2005dendritic}. Superimposed on this passive scaffold are voltage-gated sodium and calcium channels. When local depolarization exceeds threshold, these channels open and drive the membrane potential toward the reversal potential $E_r$, producing a regenerative dendritic spike that propagates toward the soma. Somatic action potentials can also back-propagate into the dendrites, reopening these channels in reverse and triggering calcium influx that enables synapse-specific plasticity \cite{bicknell2021synaptic, larkum2022dendrites}.

Among these mechanisms, N-methyl-D-aspartate (NMDA) receptors serve as time-sensitive amplifiers. Their conductance increases only when glutamate release coincides with sufficient depolarization, generating a supra-linear plateau that amplifies synchronous inputs while filtering out asynchronous or reversed spikes. In engineering terms, NMDA-rich dendrites function like temporal filters and logical AND gates. They can distinguish input sequences and timing without requiring additional network layers, supporting native temporal pattern recognition \cite{polsky2004computational, tran2015contribution, branco2010dendritic}.

Because of this compartmentalized dendritic processing, neurons implement complex signal gating through the interplay of calcium and voltage-gated channels, which respond selectively within narrow voltage windows. While a single input may generate a dendritic spike, excessive depolarization from simultaneous inputs can inactivate the channels and block the spike,  enabling XOR-like computation \cite{gidon2020dendritic}. Modeling such input-output behavior of a human pyramidal neuron required a convolutional neural network with five to eight hidden layers. Removing NMDA-based nonlinearities collapsed this to a network with just one layer \cite{beniaguev2021single}. While further complexity still exists in biology beyond that described here, these examples alone demonstrate that perceptrons are dramatic oversimplifications of real neurons.

\subsection{Key differences between BNNs and ANNs}

Understanding how BNNs' properties can inform ANNs' design requires direct comparison. Table \ref{tab:ann_bnn_difference} outlines key distinctions between the two architectures.

\begin{table}[h]
\centering
\caption{Differences between Biological and Artificial Neural Networks.}
\label{tab:ann_bnn_difference}
\begin{tabular}{|p{2.8cm}|p{4.5cm}|p{4.5cm}|}
\hline
\textbf{Property} & \textbf{BNN} & \textbf{ANN} \\
\hline
Training data requirements &
Evolutionary “pre-training”; innate threat detection and feature extraction \cite{Hubel1962ReceptiveFB,hman2001FearsPA} &
Large datasets; extensive pre-training \cite{khajehnejad2025dynamic,EgasSantander2024HeterogeneousAN} \\
\hline
Energy efficiency &
Operates on a few watts of glucose \cite{levy2020computation} &
High computational resource consumption \cite{samsi2023words} \\
\hline
Intranetwork communication &
Complex biochemical interactions at synapses \cite{chen2023overview} &
Simple mathematical operations (e.g., addition, convolution) \\
\hline
Learning mechanism &
Self-organization; synaptic plasticity; predictive coding \cite{gerstner2002mathematical,ororbia2023neuro,kilner2007predictive} &
Gradient-based backpropagation \cite{rumelhart2013backpropagation} \\
\hline
System architecture &
Biological neurons with integrated storage and computation \cite{beniaguev2021single,hari2015brain} &
Silicon-based transistors; von Neumann architecture (separate memory and processing) \cite{schuman2022opportunities} \\
\hline
Emergent properties &
Spontaneous oscillations; criticality; wave propagation; homeostatic balance \cite{kagan2025harnessing,beggs2008criticality,beggs2022cortex,heffern2021phase,weng2015brain} &
Hierarchical feature representations; attractor dynamics \\
\hline
Resilience to damage &
Dynamic rewiring and homeostatic plasticity allow partial recovery \cite{dancause2005extensive,masson2001epidemiology} &
Static redundancy (e.g., dropout); limited on-the-fly recovery \\
\hline
\end{tabular}
\end{table}

\subsection{Properties of Functional BNNs}

The internal structure of BNNs plays a central role in their capacity for learning. While modern technology enables precise fabrication of digital processors, it does not yet allow us to replicate the cellular mechanisms and synaptic interactions of BNNs. However, progress in SBI is rapidly advancing.

The following are key features required for functional, learning-capable BNNs:

\begin{itemize}
  \item \emph{Neuronal diversity:} Brain function relies on the interplay between diverse neuron types, including excitatory pyramidal cells and inhibitory interneurons, which regulate network dynamics and plasticity \cite{bittner2017population, lam2022effects, deco2014local, cadwell2019development}.
  \item \emph{Modular and hierarchical organization:} Neural circuits are organized hierarchically to process sensorimotor information and execute complex behaviors  \cite{valencia2009complex,hilger2017intelligence}. Engineered BNNs do not fully replicate this architecture, but “assembloid” and "connectoid" approaches aim to capture key aspects of modular connectivity \cite{qian2019brain,porciuncula2021age,bagley2017fused}. 
  \item \emph{Long-range communication and critical dynamics:} Oscillations rhythms such as gamma (\~30-100 Hz), theta (\~4-8 Hz), and beta (\~13-30 Hz) support inter-regional coordination of memory, attention, and motor control \cite{weiss2023neuromodulation}. These dynamics promote criticality, balancing stability and adaptability \cite{kagan2025harnessing, beggs2008criticality, beggs2022cortex, heffern2021phase, ma2019cortical}.  
  \item \emph{Parallel distributed computation:} The brain’s billions of neurons and trillions of synapses support massively parallel processing  \cite{muller2023parallel,sigman2008brain}. While modern computer architectures such as graphical processing units (GPUs) enable parallelization in ANNs \cite{keckler2011gpus}, BNNs still far exceed them in connectivity and energy efficiency.
\end{itemize}

Together, these properties highlight the importance of structural, dynamic, and cellular realism in designing synthetic systems that aim to approximate brain-like computation.

\section{Mapping the NeuroAI Spectrum}
\label{sec:neuroai_spectrum}

NeuroAI encompasses a diverse and evolving set of approaches that span biological, computational, and engineering disciplines. To provide a coherent structure for this interdisciplinary landscape, we outline a conceptual framework that organizes the field into its major components and points of intersection.

\subsection{Hardware, Software, and Wetware Domains}

Neuroscience and AI are vast, multifaceted disciplines, each encompassing a broad array of subfields. NeuroAI, as an intersection of these domains, naturally forms a diverse and expansive spectrum rather than a narrow, single-focus field. This spectrum can be conceptually factorized into three major domains \cite{kagan2025harnessing}:

\begin{enumerate}
\item \emph{Hardware:} neuromorphic chips and spike-based silicon systems
\item \emph{Software:} computational models inspired by brain function (e.g., neuro-symbolic AI, reinforcement learning)
\item \emph{Wetware:} living biological neural systems (e.g., organoids, cultures, brain slices)
\end{enumerate}

\begin{figure}
    \centering
    \includegraphics[width=1\linewidth]{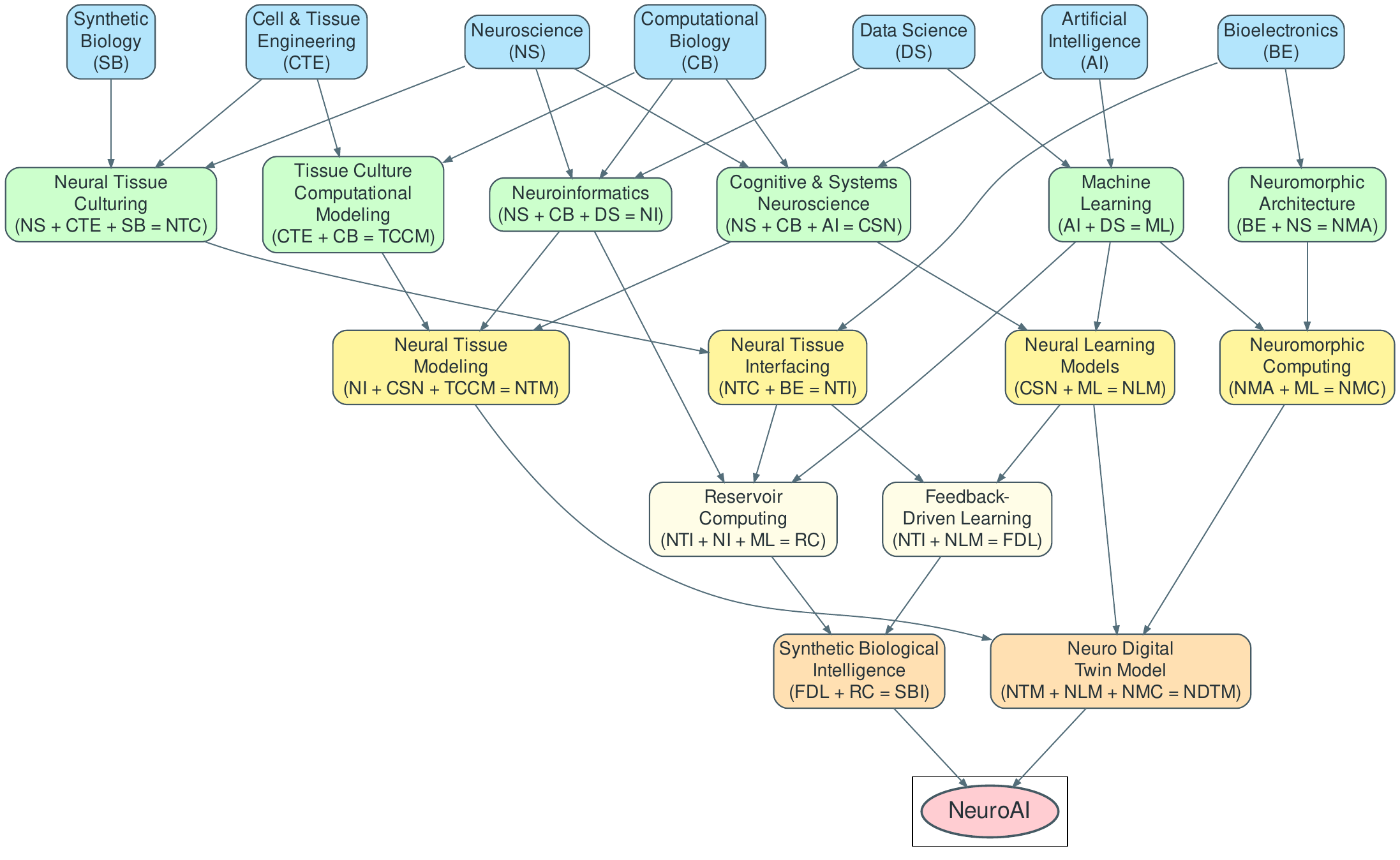}
    \caption{Graph representation of the relations between different fields that build the foundation for NeuroAI.}
    \label{fig:oi_field_graph}
\end{figure}

Figure \ref{fig:oi_field_graph} visually represents how these domains interrelate.

Table \ref{tab:oi_fields} defines the subfields and composite fields derived from these categories.

\subsection{Categorization of NeuroAI Fields}

Focusing on SBI, we categorize NeuroAI-related fields into two broad classes:

\begin{enumerate}
    \item \emph{Fields that do not involve living neural cells:} These areas leverage the morphology, physiology, and computational principles of BNNs to develop brain-inspired hardware and algorithms. This class includes:
    \begin{enumerate}
        \item \emph{Neuromorphic computing}, which seeks to emulate the connectivity and dynamics of synapses and neurons using silicon-based architectures, moving away from the conventional von Neumann architecture \cite{abdallah2022neuromorphic}.
        \item \emph{Neuro-inspired learning}, which draws from the functional strategies of biological systems to guide algorithm design for machine learning.
    \end{enumerate}

    \item \emph{Fields that use living neural cells:} Motivated by the efficiency and adaptability of BNNs, these areas integrate biological neurons with digital systems. SBI and its subsets, such as OI and more controlled Bioengineered Intelligence (BI)\cite{kagan2025two}, exemplify this approach using neural cultures to perform computational tasks. Some efforts also employ acutely harvested brain slices \cite{liu2025ex}. These BNNs can be integrated with computers to develop effective brain-computer interface (BCI) systems.
\end{enumerate}

\section{Hardware Foundations: Neuromorphic Computing}

\begin{figure}
    \centering
    \includegraphics[width=1\linewidth]{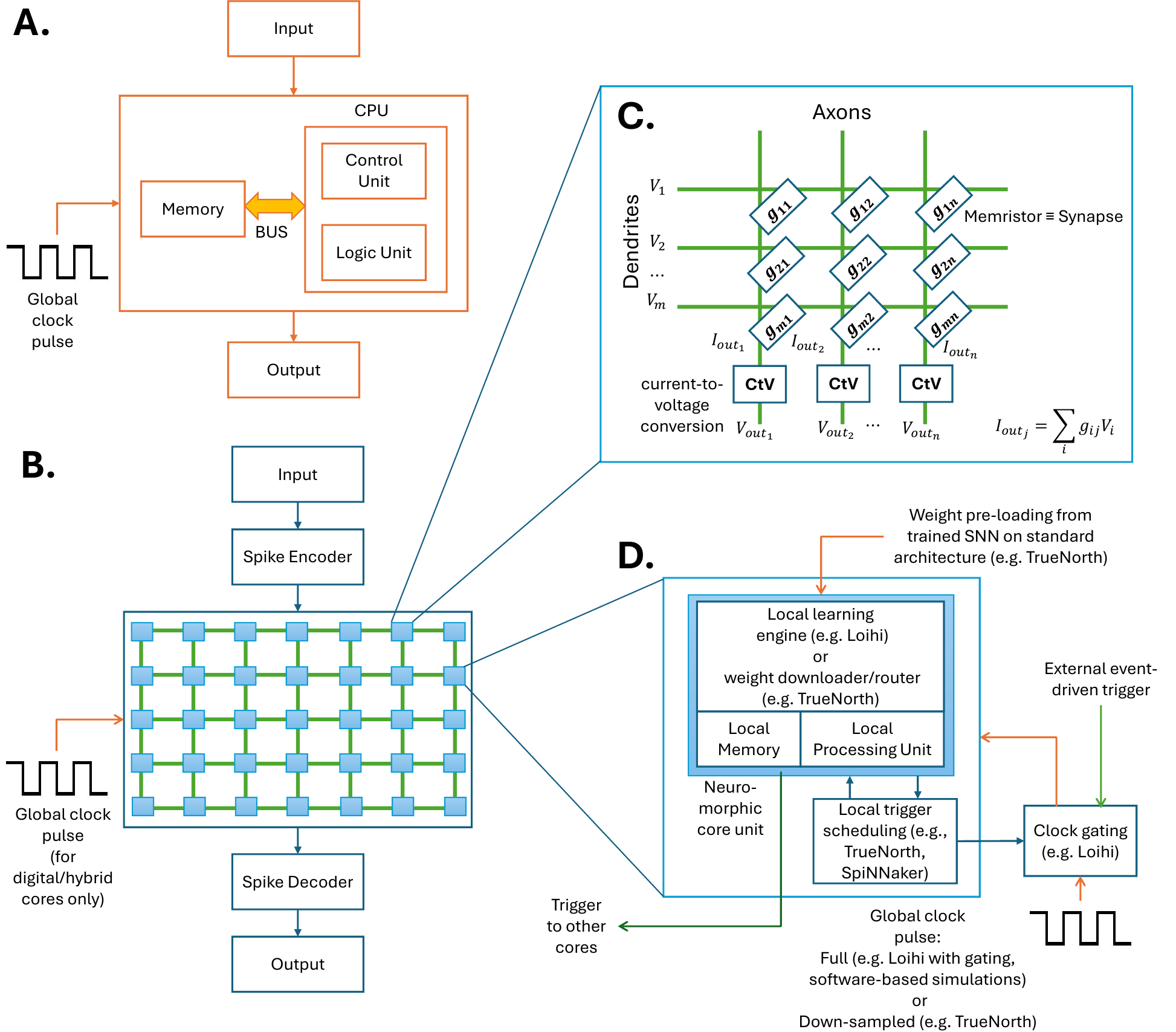}
    \caption{Illustration of the neuromorphic architecture. A: Traditional von Neumann architecture with separate memory and processing units connected by a shared communication BUS, which creates a bandwidth bottleneck and increases energy consumption. B: Neuromorphic architecture composed of analog, digital, or hybrid cores activated by spike-based signals, mimicking the event-driven dynamics of biological neurons. C: Analog neuromorphic core based on memristors-resistive elements that simultaneously store and modulate synaptic weights without requiring a clock signal. D: Digital neuromorphic core with integrated local memory and processing units. These cores can operate with pre-trained weights or perform local asynchronous updates via embedded learning engines.}
    \label{fig:neuromorphic_computing}
\end{figure}

Neuromorphic computing refers to the design of hardware systems that mimic the spike-based signaling and structural features of biological neural networks. Neuromorphic AI, in turn, implements algorithms designed for such architectures \cite{abdallah2022neuromorphic,yang2020neuromorphic,ivanov2022neuromorphic}. These systems are event-driven, activating processing elements only upon stimulation, which minimizes idle power consumption and reduces latency. Unlike traditional von Neumann architectures with separated memory and computation, neuromorphic hardware performs in-memory computation, reducing data transfer bottlenecks.

\subsection{Off-Chip and On-Chip Learning}

Neuromorphic systems support two principal learning paradigms:

\begin{enumerate}
    \item \emph{Off-chip learning and weight transfer:} Here, spiking neural networks (SNNs) are trained on standard digital hardware using deep learning frameworks augmented with surrogate gradient techniques \cite{huh2018gradient, yang2020temporal}. The trained weights are then deployed to neuromorphic devices for efficient inference. Examples include IBM’s TrueNorth \cite{akopyan2015truenorth} and SpiNNaker \cite{furber2012overview}, which support large-scale simulation of SNNs.
    \item \emph{On-chip local learning:} In contrast, some neuromorphic devices support local, real-time learning using biologically inspired rules such as spike-timing-dependent plasticity (STDP) and Hebbian learning \cite{yang2020neuromorphic,kempter1999hebbian}. Intel’s Loihi, for example, uses embedded learning engines to perform asynchronous weight updates \cite{davies2018loihi}. Analog systems using memristors or silver nanowire networks adjust conductance in response to local voltage or current patterns \cite{loeffler2023neuromorphic, qiao2017analog, park2022experimental}.
\end{enumerate}

Figure \ref{fig:neuromorphic_computing} summarizes these architectural components. A comparative overview of major neuromorphic platforms is in Table S2. While neuromorphic hardware currently outperforms wetware in terms of scalability and robustness, limitations in component reliability (especially for memristors) remain a bottleneck \cite{ye2022overview,wang2024can}. Nevertheless, the asynchronous and event-driven nature of neuromorphic systems makes them promising candidates for hybrid integration with BNNs \cite{kagan2023technology,kagan2025harnessing,smirnova2023organoid}.

\section{Software Models: Neuro-Inspired Learning}

Neuro-inspired learning seeks to emulate how biological systems acquire and adapt knowledge. While modern deep neural networks (DNNs) are powerful in recognizing patterns, they often lack flexibility in extrapolation, adaptation, and long-term reasoning: capacities where biological intelligence excels. By drawing on principles from neuroscience, neuro-inspired models develop more robust, efficient, and generalizable learning systems. These efforts represent a key component of the NeuroAI spectrum.

\subsection{Neuro-symbolic AI and Agentic AI}

Neuro-symbolic AI integrates the strengths of symbolic reasoning (explicit rules, logical inference) with sub-symbolic learning (e.g., deep learning for perception and pattern recognition) \cite{sheth2023neurosymbolic, garcez2023neurosymbolic}. This hybrid approach enables models to combine structured knowledge with data-driven inference, enhancing accuracy, explainability, and adaptability.

The training of Neuro-symbolic AI differs from popular AI models, as the former incorporates the following important aspects:

\begin{itemize}
    \item \emph{Knowledge graph (KGs):} Graph-based structures that encode entities and relationships, enabling models to retrieve and reason over domain knowledge  \cite{hogan2021knowledge, ammanabrolu2021learning, reinanda2020knowledge}. When combined with large language models (LLMs) via retrieval-augmented generation (RAG), KGs improve factual grounding and reduce hallucinations \cite{lewis2020retrieval}. 
    \item \emph{Symbolic reasoning and rules:} Logical inference engines operate on graph-based facts to produce transparent, traceable decisions. This contrasts with black-box neural reasoning. Frameworks like DeepProbLog \cite{manhaeve2018deepproblog} and Logic Tensor Networks \cite{serafini2016logic} exemplify this integration, enabling models to follow formal logic while learning from data \cite{yi2018neural,mao2019neuro,han2019visual,trinh2024solving}.

\begin{figure}
    \centering
    \includegraphics[width=1\linewidth]{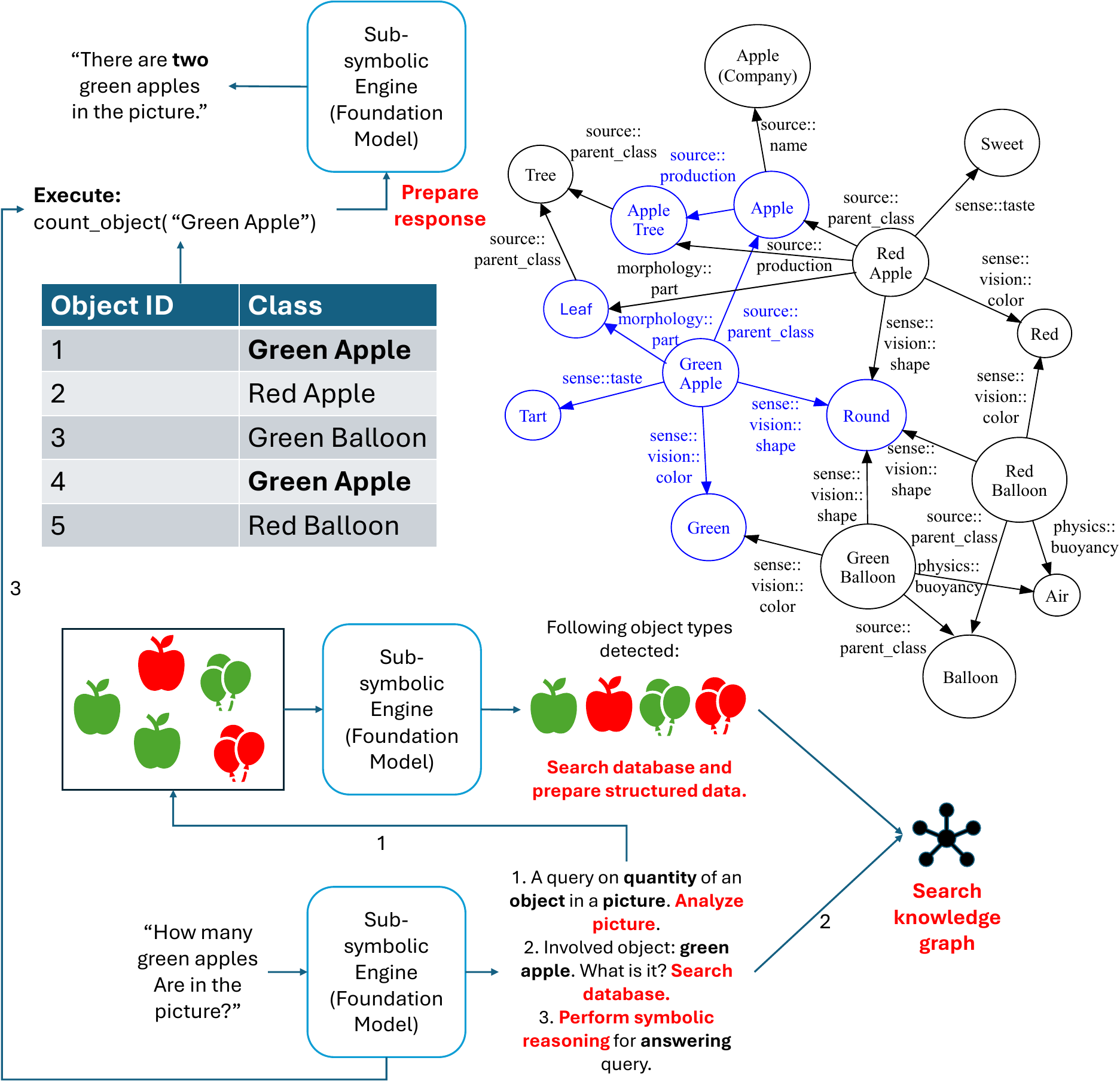}
    \caption{Exemplary neuro-symbolic AI model. A multimodal foundation model (sub-symbolic core) receives a user query, activates a relevant subgraph from an internal knowledge graph, performs symbolic reasoning, and generates an output. This hybrid architecture improves factual reliability and interpretability over purely sub-symbolic systems.}

    \label{fig:neurosymbolic_ai}
\end{figure}

    \item \emph{Transparency and ethics:} Symbolic layers enhance interpretability and support the incorporation of ethical constraints, particularly important for agentic systems that interact with dynamic environments \cite{dingli2023neuro,lu2024surveying}.

    \item \emph{Multimodal data:} Neuro-symbolic frameworks can incorporate diverse data modalities (text, images, audio) into coherent knowledge representations, supporting complex tasks such as medical reasoning \cite{khan2024neusyre,niu2025medical,niu2024development}.

    \item \emph{Agentic AI:}  Neuro-symbolic agents can be made to autonomously interact with their environment, update their internal knowledge representations, and adapt rules through active learning mechanisms \cite{mendonca2021discovering,du2023can}. These systems are often grounded in active inference principles \cite{friston2016active}, enabling context-sensitive, goal-directed behavior.

\end{itemize}

Figure~\ref{fig:neurosymbolic_ai} illustrates how these hybrid systems operate. Neuro-symbolic AI is especially well-suited for interfacing with complex BNNs such as \textit{in vitro} neural cultures. Such systems can interpret human input, decode neural signals, and modulate feedback stimulation with contextual awareness, potentially acting as interpreters, mediators, co-learners, and co-workers in human-BNN hybrid systems.

\subsection{Reinforcement Learning and Active Inference}

For task-oriented objectives, intelligent systems must be capable of perceiving their environment, making decisions, and adapting behavior based on feedback. Two leading computational frameworks have emerged to formalize these processes: reinforcement learning (RL) and active inference (AIF).

\emph{Reinforcement learning} is a model of decision-making in which an agent interacts with the environment by observing states ($s$), taking actions ($a$), and receiving scalar rewards ($R$). The agent's goal is to learn a policy that maximizes cumulative reward using methods such as Q-learning and actor-critic \cite{wiering2012reinforcement, chung1967markov}. State observations ($o$) may be full or partial, and value functions like \( V(s) \) or the state-action value function \( Q(s, a) \) guide decision-making through recursive relationships such as the Bellman equation:
\begin{equation}
    V(s_t) = \mathbb{E}_{a_t, s_{t+1}} \left[ R(s_t, a_t) + \gamma V(s_{t+1}) \right]
\end{equation}
Where $\gamma$ is a discount factor. Because RL optimizes for reward, it does not include an intrinsic mechanism for exploration. Exploration strategies must be added externally, such as $\epsilon$-greedy action selection or stochastic policies.

\begin{figure}
    \centering
    \includegraphics[width=1\linewidth]{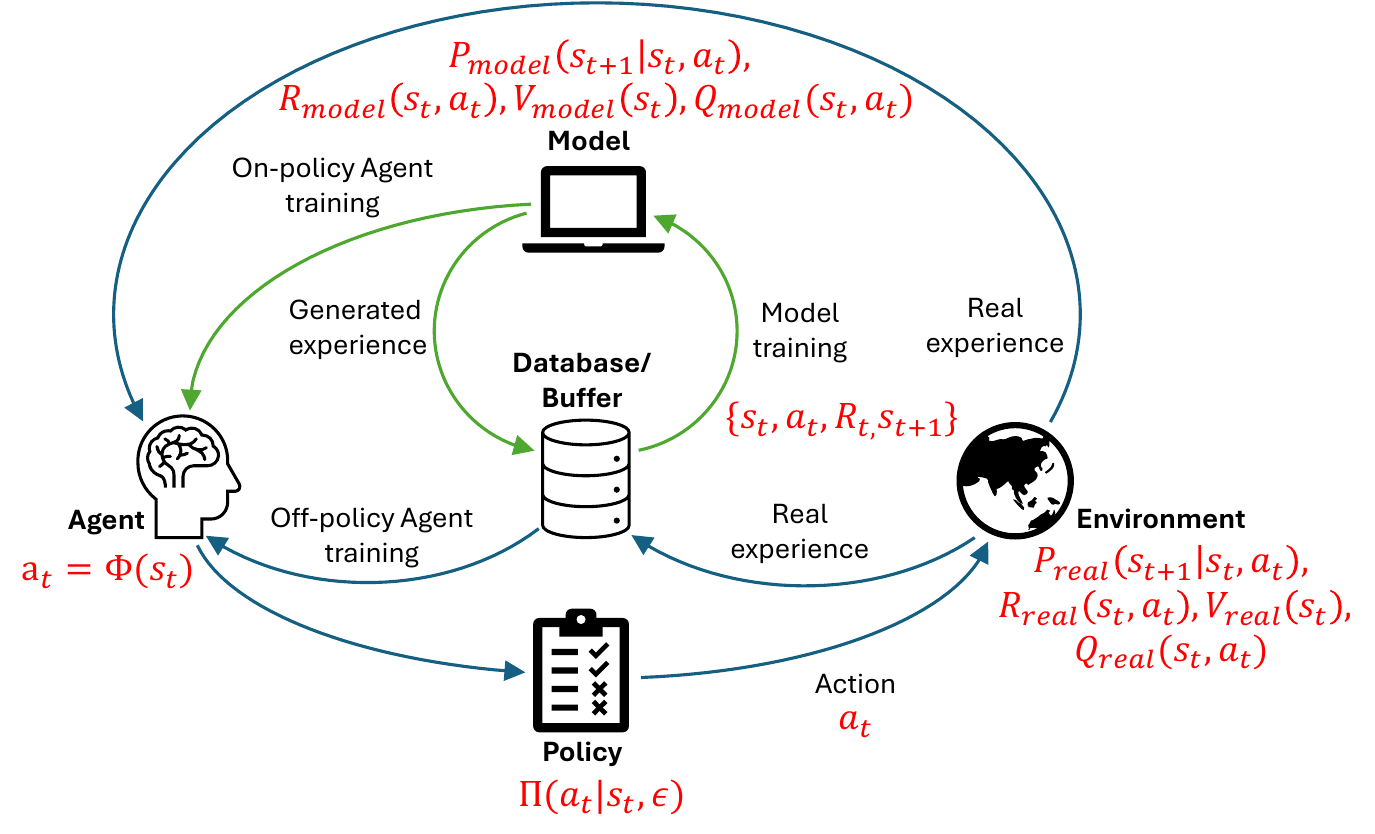}
    \caption{Diagram of reinforcement learning. Green arrows indicate a model-based approach. Experience is stored in an offline buffer to enable off-policy learning or used directly for on-policy updates.}

    \label{fig:reinforcement_learning}
\end{figure}

RL may be model-free, relying solely on interactions, or model-based, where an internal simulation of the environment supports policy evaluation and planning. Figure~\ref{fig:reinforcement_learning} illustrates these architectures. Deep RL extends this framework using artificial neural networks to approximate value functions, making it compatible with modern machine learning infrastructure.
In the context of SBI, frameworks similar to RL have been used to associate external rewards with biochemical or electrical stimulation in biological neural networks. However, calibrating appropriate reward signals and decoding internal states remain active topics.

\emph{Active inference}, in contrast, is rooted in the Free Energy Principle (FEP) \cite{friston2010free, friston2016active}, which posits that agents minimize a variational bound on surprise. Surprise is defined as the negative log-likelihood of sensory input under the agent’s internal model and reflects the discrepancy between actual and expected observations. Active inference assumes that agents do not observe the true hidden state $s^*$ directly, but instead receive partial observations $o$ and form an internal belief about hidden causes, denoted by $s$. The agent maintains an internal generative model represented as the joint distribution \( p(s, o) \). Perception involves minimizing the discrepancy between the likelihood \( p(o \mid s) \) and the marginal likelihood \( p(o) \), leading to Bayesian belief updating:

\begin{equation}
    p(s_t \mid o_t) = \dfrac{p(o_t \mid s_t)\,p(s_t)}{p(o_t)}
\end{equation}

where $t$ is the current time instant. However, the marginal likelihood \( p(o_t) = \int p(o_t, s_t)\, ds_t \) is often intractable. To overcome this, active inference introduces a variational approximation \( q(s_t) \approx p(s_t \mid o_t) \) and minimizes the KL divergence \( D_{\mathrm{KL}}[q(s_t) \,\|\, p(s_t \mid o_t)] \), which is equivalent to minimizing the variational free energy:

\begin{equation}
    F(o_t) = D_{\mathrm{KL}}[q(s_t) \,\|\, p(s_t \mid o_t)] - \log p(o_t)
\end{equation}

This formulation captures the perceptual (or passive) aspect of inference at time $t$. However, active inference also includes the agent's ability to take actions \( a_t\), following a policy \( \pi = \{a_t, a_{t+1}, \cdots a_T\}\) that influence future hidden states \( s^*_{t+1} \) and observations \( o_{t+1} \). Thus, it is necessary to minimize not only the current variational free energy \( F(o_t) \), but also the total expected free energy \( G_{t:T}(\pi) \) over a future policy \( \pi \), at a future time step\( \tau \):

\begin{equation}
    G_{\tau}(\pi) = \mathbb{E}_{q(o_\tau, s_\tau \mid \pi)} \left[ \log q(s_\tau \mid \pi) - \log p(o_\tau, s_\tau) \right] \\
\end{equation}

The optimal policy is then defined by the joint minimization of the current and expected free energy:

\begin{equation}
    \pi^* = \arg\min_{\pi} \left[ F(o_t) + \sum_{\tau = t+1}^{T} G_{\tau}(\pi) \right]
\end{equation}

\begin{figure}
    \centering
    \includegraphics[width=1\linewidth]{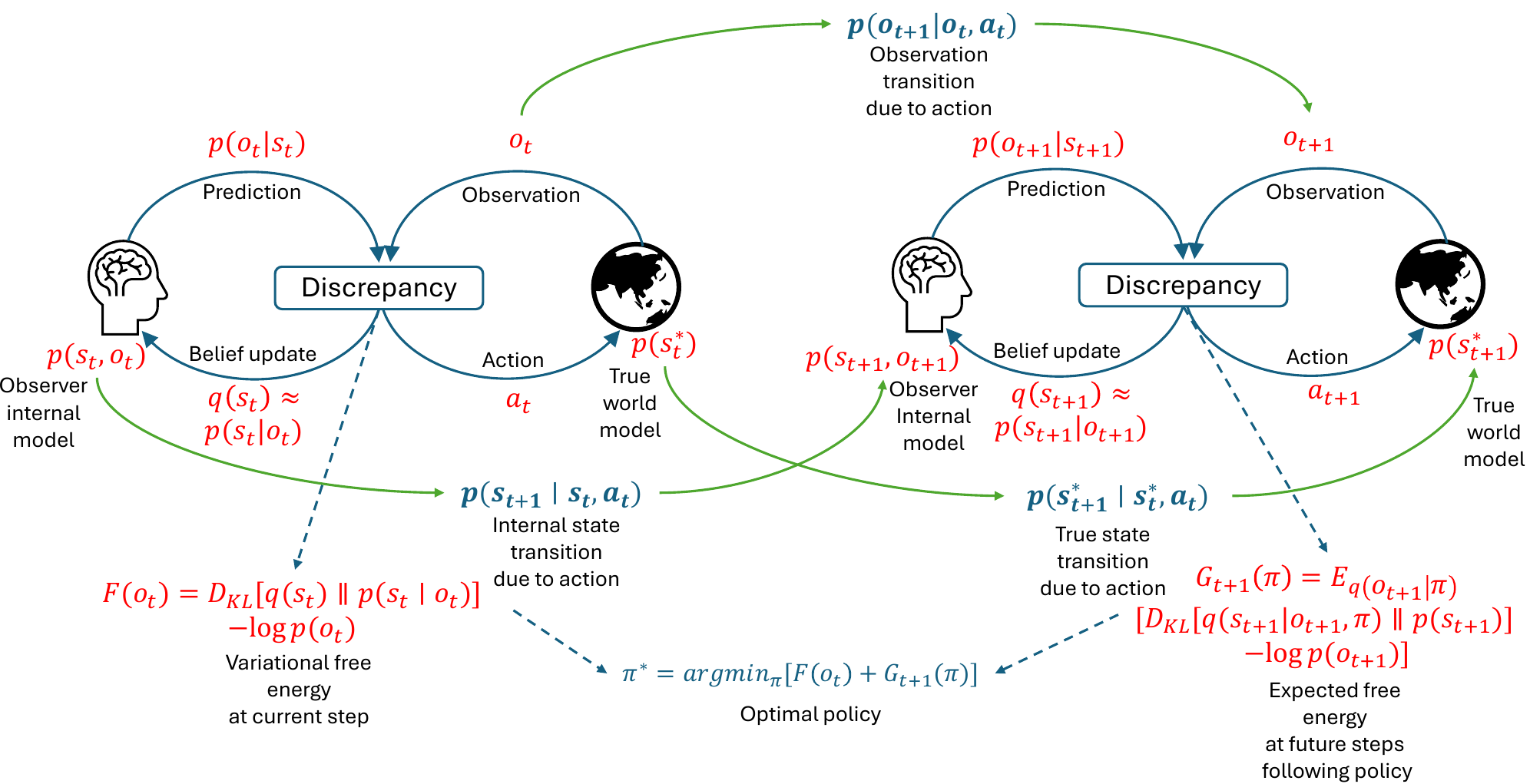}
    \caption{Diagram of active inference over two timesteps. The agent updates internal beliefs about hidden causes and selects actions that minimize both current and expected free energy.}
    \label{fig:active_inference}
\end{figure}

Figure \ref{fig:active_inference} illustrates how active inference provides an optimal policy by minimizing the discrepancy of the internal model and the observed environment. The expected free energy can be decomposed into intrinsic (epistemic) and extrinsic (goal-directed) components:

\begin{equation}
    G_\tau(\pi) = \underbrace{\mathbb{E}_{q(o_\tau \mid \pi)}[-\log p(o_\tau)]}_{\text{extrinsic (risk)}} + \underbrace{\mathbb{E}_{q(o_\tau \mid \pi)} \left[ D_{\mathrm{KL}}\left(q(s_\tau \mid o_\tau, \pi) \,\|\, p(s_\tau)\right) \right]}_{\text{intrinsic (information gain)}}
\end{equation}
Extrinsic risk captures the anticipated surprise or negative consequences of future observations, quantified by how improbable or undesirable those outcomes are under the agent’s current model. Intrinsic information gain measures the expected reduction in uncertainty about hidden states, reflecting how much a new observation is likely to improve the agent’s internal beliefs. This decomposition illustrates why exploration is inherent to AIF: agents are naturally driven to seek information that reduces uncertainty about their environment. In contrast to RL, AIF requires agents to update a generative model of the environment, which guides both perception and action. 

To implement AIF in deep learning systems, neural networks are used to approximate generative models of sensory input and latent causes \cite{friston2018deep,fountas2020deep,shi2024blind}. In SBI applications, feedback signals can be adapted to drive BNNs toward low-prediction-error states \cite{kagan2022vitro}, enabling goal-directed behavior in neural cultures \cite{robbins2024goal}.

Although RL and AIF differ in both conceptualization and formulation, they do share similarities in some aspects. Reward maximization in RL can be viewed as minimizing expected surprise, and both can use internal models to evaluate actions before execution. Table~\ref{tab:reinforcement_activeinference} summarizes the key contrasts between the two approaches.

\begin{table}[h]
\centering
\caption{Comparison of Reinforcement Learning and Active Inference.}
\label{tab:reinforcement_activeinference}
\begin{tabular}{|p{1.5cm}|p{5.25cm}|p{5.25cm}|}
\hline
\textbf{Aspect} & \textbf{Reinforcement Learning (RL)} & \textbf{Active Inference (AIF)} \\
\hline
Core objective &
Maximize expected cumulative reward provided by the environment. &
Minimize variational free energy (upper bound on surprise) by aligning predictions with observations. \\
\hline
Learning mechanism &
Temporal-difference learning or policy gradients using reward prediction errors. &
Variational Bayesian inference by minimizing free energy or prediction error. \\
\hline
Action selection &
Argmax over value functions \( V(s) \) or \( Q(s, a) \); does not explicitly model hidden states or future outcomes. &
Policy inference via minimization of expected free energy under a generative model. \\
\hline
Exploration &
Driven by heuristics (e.g., $\epsilon$-greedy, UCB) or stochastic policies. &
Emerges naturally from epistemic value (uncertainty reduction) in expected free energy. \\
\hline
Model requirement &
Model-free and model-based variants exist; internal model optional. &
Requires an internal generative model of states and observations. \\
\hline
ANN integration &
Widely supported (e.g., deep Q-networks, policy gradients) via standard ML libraries. &
Implemented using deep generative models (e.g., deep VAEs) with custom variational inference. \\
\hline
BNN applicability &
Maps rewards to stimuli (e.g., electrical or chemical), but difficult to calibrate feedback. &
Can guide self-organizing dynamics through feedback that implicitly modifies prediction error. \\
\hline
\end{tabular}
\end{table}

While neuro-symbolic architectures, RL, and AIF are commonly applied to purely artificial systems, these models are increasingly relevant in the context of biological substrates. In particular, hybrid systems that combine living neural cultures with digital controllers rely on software models to decode neural activity, interpret internal states, and guide feedback stimulation. As such, the learning algorithms described above are not just theoretical frameworks but active components in emerging SBI platforms. It is important to note, however, that current cell culturing methods place constraints on whether RL or AIF is best applied. Without a discrete reward function being accessible, RL can be a challenging approach \cite{friston2025active}. Yet while AIF may be used in the absence of this control, it may lead to difficulty in interpreting or falsifying what biology mechanism of action drives any observed learning. Both these areas require significantly further development \cite{kagan2025harnessing}. In the next section, we examine how biological and artificial elements are integrated for brain-computer interfaces and synthetic biological intelligence systems.

\section{SBI Systems}

BCIs translate neural activity into machine-interpretable signals and vice versa, enabling the study, augmentation, or restoration of cognitive and motor functions \cite{shih2012brain}. While traditional BCI research has focused on clinical applications such as neuroprosthetics and rehabilitation, emerging biohybrid approaches are expanding this framework toward cognitive augmentation and new computational paradigms. By uniting plastic, self-organizing BNNs with fast and deterministic digital processors, BCI enables open-loop and closed-loop systems that restore movement, enhance perception, and accelerate problem-solving \cite{cinel2019neurotechnologies}. 

SBI extends this integration by constructing hybrid systems that incorporate living neural cells as computational substrates. These efforts are motivated by the energy efficiency, self-organization, and adaptability of BNNs \cite{kagan2023technology}. Since interfacing with the brain \textit{in vivo} presents surgical, immunological, and ethical challenges, researchers have turned to \textit{in vitro} systems, including neural cultures and organoids, to implement experimental models of intelligence \cite{waisberg2024ethical, kagan2024embodied}.

\subsection{Open-loop vs Closed-loop Learning}

Two major approaches have been developed for using \textit{in vitro} neural cultures in computation:

\begin{figure}
    \centering
    \includegraphics[width=1\linewidth]{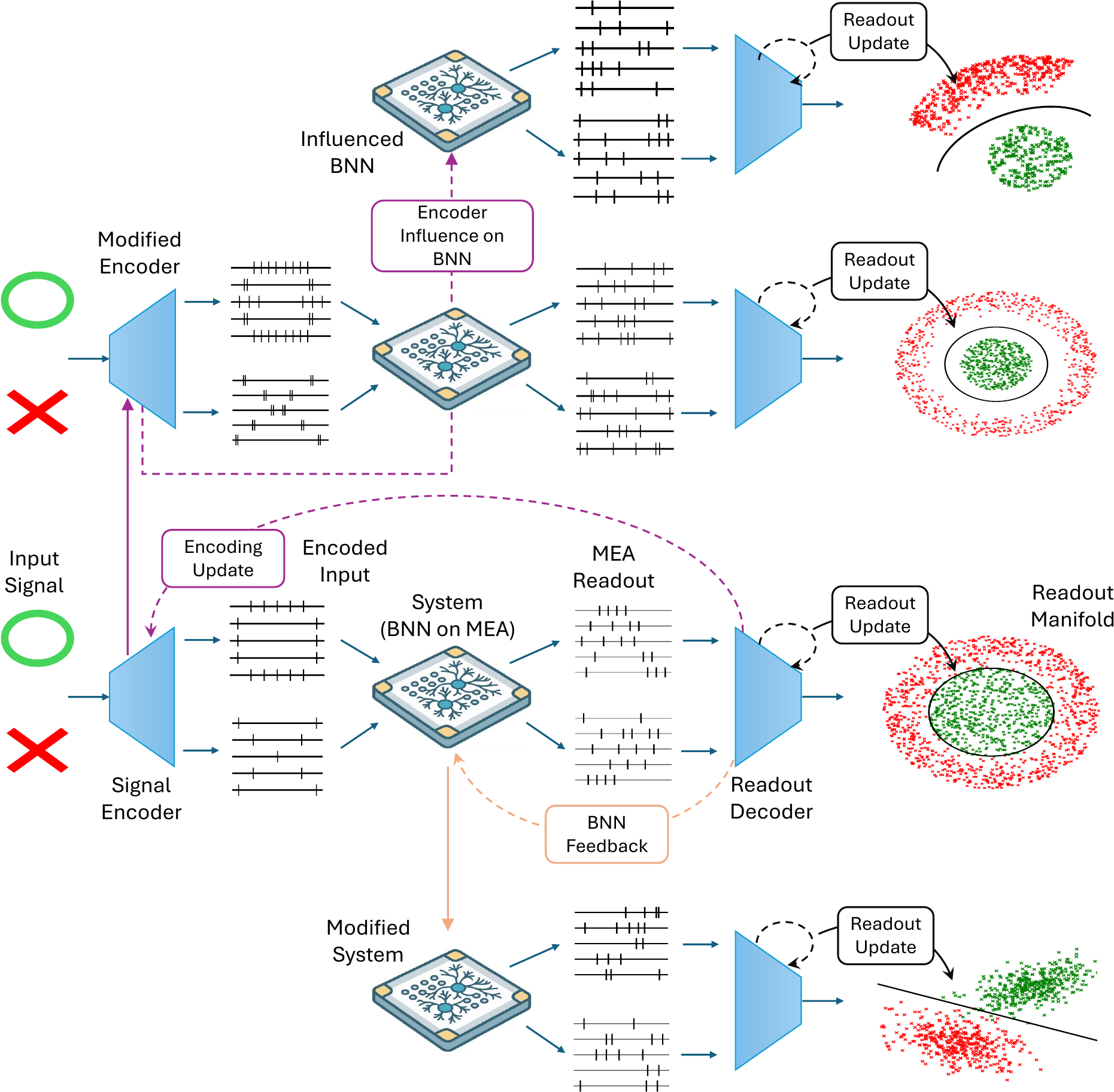}
    \caption{SBI computing methods. As an example, we only show the separation of two signals here. The computing method relies on open-loop configurations such as reservoir computing, where only the readout layer is updated to separate the two signals. Closed-loop configuration allows modifying the signal encoder or directly influencing the BNN through electrical/chemical/mechanical/optical means, regulated by a signal comparator or complex task-oriented programming. It must be noted that as a dynamic and self-organizing system, even changes in encoder patterns can have an influence on the BNN itself.}
    \label{fig:organoid_computing_methods}
\end{figure}

\begin{enumerate}

    \item \emph{Reservoir computing (RC):} This approach treats the BNN as a high-dimensional dynamical system that transforms input signals into a rich spatiotemporal response. The synaptic connections of the BNN are not explicitly trained; instead, a separate readout layer is trained to map neural activity to desired outputs \cite{smirnova2023reservoir}. RC leverages the natural nonlinear dynamics and fading memory of biological systems for tasks such as pattern recognition \cite{sumi2023biological}, chaotic time series prediction \cite{lindell2024chaotic}, and blind source separation \cite{isomura2015cultured}. 
    
    Although reservoir systems are often assumed to have fixed weights, BNNs exhibit synaptic plasticity \cite{caporale2008spike,mostajo-radji_fate_2024} and may self-organize based on input-driven plastic changes. This can enhance performance even in nominally open-loop configurations.
    
    \item \emph{Feedback-based learning:} In contrast to RC, closed-loop systems continuously adapt neural activity through real-time feedback. External stimulation is used to reinforce or inhibit specific network patterns, allowing the BNN to modify its synaptic architecture through mechanisms such as STDP or Hebbian learning \cite{caporale2008spike,kempter1999hebbian,gerstner2002mathematical}. This enables the BNN to develop goal-directed behavior, often framed in terms of active inference \cite{friston2016active, voitiuk2024feedback,robbins2024goal}. Unlike passive computation, these systems support iterative improvement. A notable example is DishBrain, a system in which neural cultures learn to play a simplified video game using feedback signals \cite{kagan2022vitro}.
\end{enumerate}
Figure \ref{fig:organoid_computing_methods} contrasts these two paradigms. While RC offers efficiency and simplicity, closed-loop learning provides greater flexibility and higher potential for dynamic tasks.

\subsection{Modular Architectures: Assembloids and Connectoids}

Standalone brain organoids or homogeneous neural cultures often lack the structural and functional complexity necessary to model higher-order cognitive functions. These simplified systems typically do not form laminar architectures, develop long-range connectivity, or include essential non-neuronal components such as glial cells, vasculature, or immune cells \cite{lancaster2014generation}. As a result, they remain immature and limited in their ability to emulate coordinated neural dynamics.

In contrast, the mammalian brain is composed of interconnected modules, each with specialized cellular composition, connectivity motifs, and molecular identities. These modules support distinct functions such as sensory integration, memory, language, and autonomic regulation, while communicating through well-defined pathways \cite{swanson2024neural,bertolero2015modular}. This modular organization underlies both robustness and computational power. To replicate these properties \textit{in vitro}, SBI has adopted modular platforms that assemble multiple organoid components into unified systems.

Three leading approaches are:

\begin{enumerate}
  \item \emph{Assembloids (fused organoids):} In this approach, region-specific organoids are grown separately and then physically fused. The resulting structure allows spontaneous axonal projections and cell migration across regions, enabling the formation of inter-regional circuits. Assembloids preserve many of the self-organizing features of brain development and are well suited for studying interneuron migration, circuit formation, and inter-region signaling \cite{bagley2017fused,makrygianni2021brain}.

  \item \emph{Connectoids (microchannel-linked organoids):} Here, individual organoids are maintained in separate microfluidic chambers and connected via engineered microchannels that guide axonal growth between compartments. This architecture offers precise control over connectivity, environmental conditions, and recording access to each module. This approach captures some physiological relevance, but with increased control in connectivity between neural organoids \cite{osaki2024complex}.
  
  \item \emph{Bioengineered Intelligence:} An alternative approach that aims to eschew macro-scale physiological relevance has recently been proposed \cite{kagan2025two}. This method may culture neural cells as isolated networks that lack any form of macro-physiological relevance and instead represent discrete circuits with highly specified properties and functionality \cite{murota2025precision, forro2018modular}. These architectures have previously been shown to allow these relatively simple, yet highly structured, organizations to act in highly complex ways \cite{sumi2023biological, tessadori2012modular, girardin2022topologically, habibey2024engineered},      

\end{enumerate}

These different modular systems form the basis of a continuum that spans from those that more faithfully model macro \textit{in vivo}-like structure and function to those with only circuit-level similarities. Whether modeling \textit{in vivo} architectures is desirable depends purely on the research question posed. Regardless, by having a variety of approaches for controlling inter-region communication, region-specific maturation, and the inclusion of non-neuronal cell types, these modular SBI approaches provide a range of foundations for modeling development, disease, therapy, and emergent intelligence.

\section{Stability, Plasticity, and Learning Over Time}

Learning systems face a stability–plasticity dilemma, where they must encode new information without erasing prior knowledge. In the brain, this challenge is addressed by multi-timescale synaptic mechanisms. Short-term plasticity, driven by vesicle depletion and calcium dynamics, supports rapid and reversible adaptation \cite{zucker2002short}, while NMDA-receptor-dependent long-term potentiation and depression embed lasting changes when inputs are sufficiently salient \cite{bliss1993synaptic}.

At the synaptic level, active synapses form molecular tags that capture newly synthesized proteins and consolidate salient events into lasting engrams \cite{frey1997synaptic}. Metaplasticity modulates the thresholds for potentiation and depression based on prior activity, helping prevent runaway plasticity \cite{abraham1996metaplasticity}. These processes lead to sparse engram cell ensembles whose coordinated replay during rest and sleep reinforces memory traces without disrupting unrelated circuits \cite{josselyn2015finding}. At the network level, homeostatic scaling adjusts synaptic gains to stabilize overall firing rates \cite{turrigiano2004homeostatic}, while structural plasticity, such as dendritic spine remodeling and circuit rewiring, encodes information in more persistent anatomical forms \cite{fiala2002dendritic}.

In contrast, ANNs trained sequentially through backpropagation often suffer from catastrophic forgetting, where learning new tasks overwrites important parameters from earlier ones. Several approaches have been developed to mitigate this limitation. Elastic Weight Consolidation penalizes changes to weights deemed essential for previous tasks \cite{kirkpatrick2017overcoming}, while Synaptic Intelligence tracks the importance of parameters during training \cite{zenke2017continual}. In parallel, experience replay reintroduces examples from earlier tasks \cite{lopez2017gradient}, and generative replay synthesizes past-like inputs using learned models \cite{shin2017continual}. Architectural strategies, such as Progressive Networks \cite{rusu2016progressive} and dynamically expandable structures \cite{yoon2017lifelong}, allow networks to grow and accommodate new knowledge without interference.

Biological feedback loops support working memory and contextual processing, functions that are reflected in recurrent neural networks such as LSTMs and GRUs. These models retain relevant information across long sequences, reducing the impact of temporal gaps \cite{hochreiter1997long, cho2014learning}. Memory-augmented architectures, including Neural Turing Machines and Memory Networks, provide external memory buffers for selective retrieval of stored information, a strategy reminiscent of engram reactivation in biological systems \cite{graves2014neural, sukhbaatar2015end}.

Altogether, these methods parallel how the brain combines synaptic tagging, homeostatic regulation, structural adaptation, and ensemble-level replay to support learning over time. By layering mechanisms across temporal and organizational scales, both biological and artificial systems can remain adaptable while preserving accumulated knowledge.

\section{Digital Twins and the Role of Big Data in NeuroAI}

In recent years, data-driven modeling has become increasingly central to understanding complex systems that are analytically or empirically intractable. BNNs, with their dynamic, nonlinear, and evolving properties, are especially suited to this approach. Within the field of NeuroAI, digital twins offer a powerful framework for modeling and analyzing BNNs \textit{in silico}. These models can support synthetic brain-computer integration, serve as predictive tools for simulating perturbations, and reduce the need for repeated culturing of fragile living tissues.

\subsection{Digital Twin Models}

A digital twin is an accurate and continuously updated virtual model of a physical entity. Digital twins of BNNs must capture dynamic processes in real-time, enabling simulation, diagnosis, and decision-making based on live biological data. This capability is particularly important for organoid and culture systems, which are susceptible to environmental fluctuations, contamination, and aging. Digital twin models can incorporate multiple layers of biological representation, including morphological growth, electrophysiological activity, mechanochemical interaction in the extracellular matrix, gene and biomarker expression, energy metabolism, and predictions of stress responses, apoptosis, or drug sensitivity \cite{wang2022development, moller2021digital, de2025challenges}.

\begin{figure}
    \centering
    \includegraphics[width=1\linewidth]{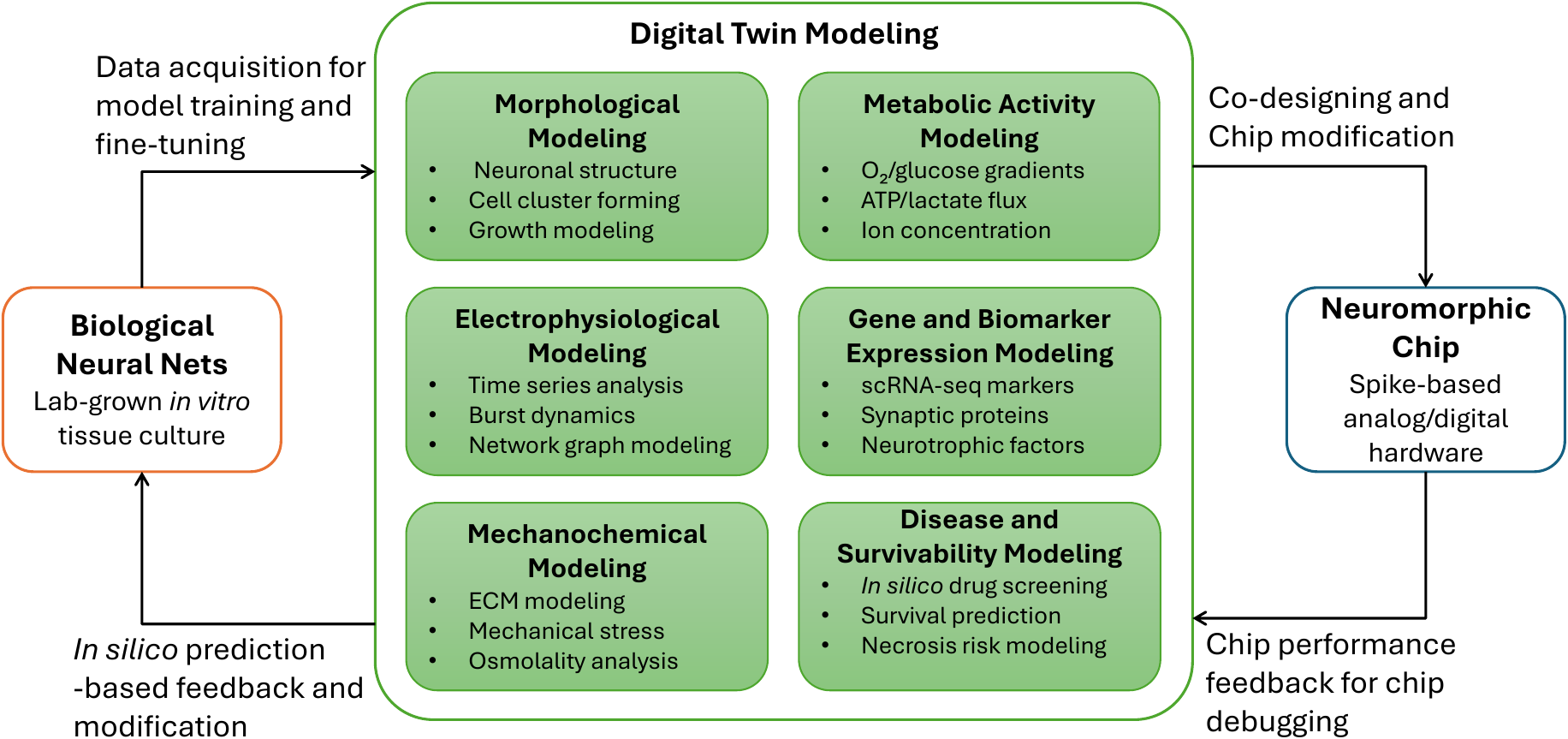}
    \caption{Diagram illustrating how BNNs, neuromorphic chips, and digital twin models (which could be based on neuro-symbolic AI) complement each other.}
    \label{fig:digital_twin}
\end{figure}

As shown in Figure \ref{fig:digital_twin}, digital twins can guide the development of neuromorphic chips by mapping activity and architecture from BNNs into hardware design. Data collected from living networks supports training and refinement of these virtual models. Neuro-symbolic AI methods, in particular, allow for interpretable reasoning over structured biological systems and help bridge learned representations with logical inference. However, building such models requires large and diverse datasets with standardized formats. Fortunately, ongoing efforts in multimodal data generation and sharing are making this feasible \cite{sharf2022functional,el2024human,andrews2024multimodal,elliott2024pathological,geng2024multiscale}. These datasets can also be used to develop foundation models that learn across spatiotemporal scales and biological modalities \cite{wang2023brainbert, wang2025foundation}, enabling unified digital representations of BNNs.

\subsection{Spike Sorting and Data Pipelines}

Electrophysiological recordings from thousands of neurons are now possible with high-density MEAs. For SBI systems and digital twins, understanding spiking dynamics is essential. However, spike sorting large datasets remains challenging. Manual sorting is no longer scalable, leading to the development of automated pipelines that use statistical and computational techniques to isolate spikes and identify single units \cite{paulk2022large}.

These pipelines begin with raw voltage traces acquired through systems such as MaxWell Biosystems or Neuropixels \cite{siegle2017open, jones2023building, paulk2022large, andrews2024multimodal, williams2022alzheimer, kagan2022vitro}. Before spike sorting can be performed, the data undergoes a series of preprocessing steps. This pipeline generally includes band-pass filtering (usually between 300 and 6000 Hz) to isolate action potentials from background noise and Local Field Potentials, followed by common-average referencing to reduce shared noise across channels \cite{buccino2022spike}. Whitening is then applied to decorrelate the signals, enhancing the separability of neural sources. In some cases, drift correction is performed to compensate for the gradual movement of neurons relative to the recording electrodes \cite{pachitariu2024spike,van2024tracking}. These preprocessing steps are often implemented within integrated frameworks or pipelines \cite{buccino2020spikeinterface, geng2024multiscale}.

\begin{figure}
    \centering
    \includegraphics[width=1\linewidth]{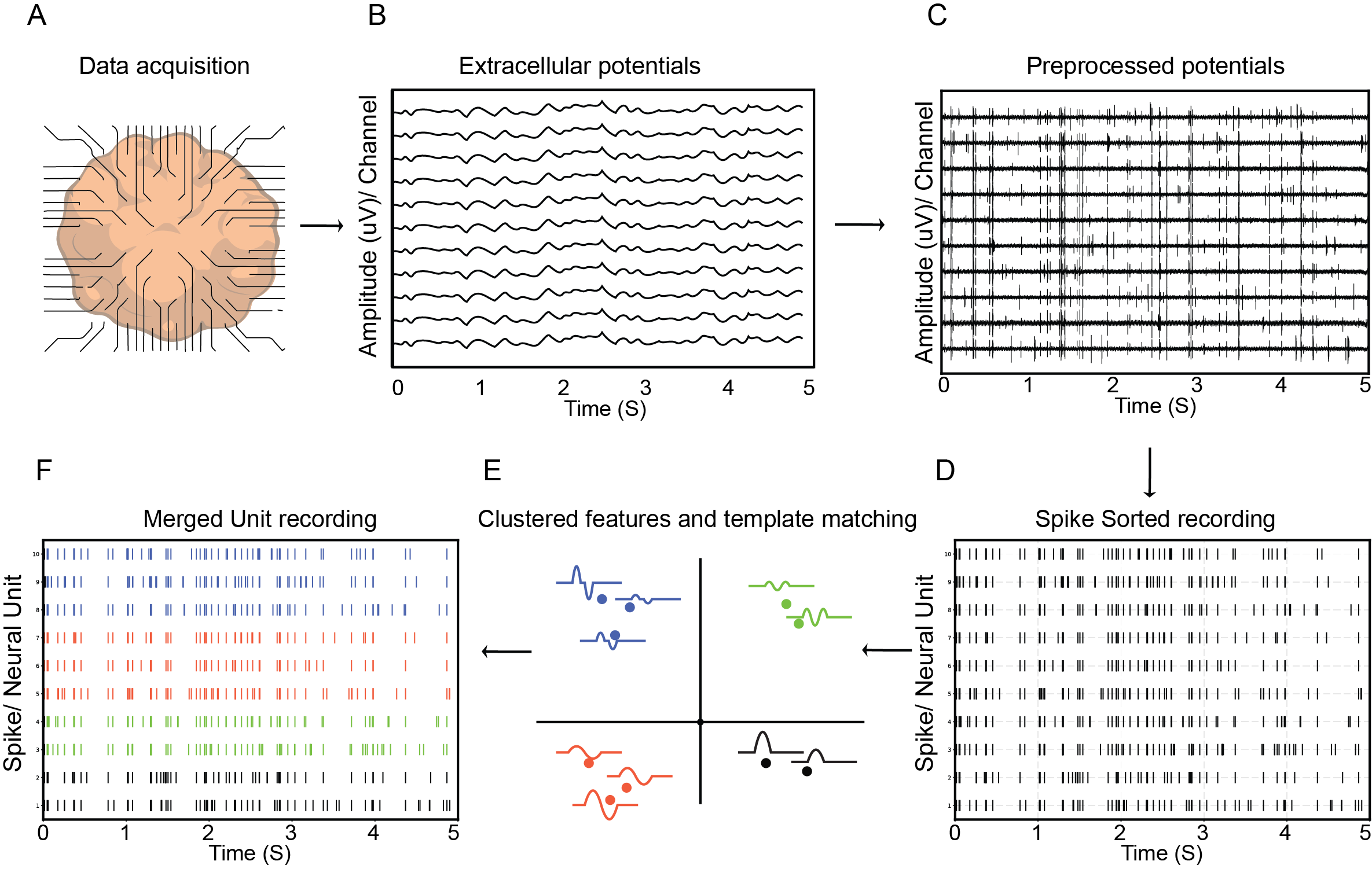}
    \caption{Representation of a typical spike sorting pipeline. A) Data acquisition from a BNN. B)Extracellular potentials are obtained from the recording. C) Raw potentials undergo processing steps like whitening, denoising, and band pass filtering. D) Spikes are detected using thresholding. E) Features are extracted and templates matched. F) Units are aggregated across channels.}
    \label{fig:spikeSortingPipeline}
\end{figure}

After preprocessing, candidate spikes are detected and aligned into waveform windows. These are projected into a feature space, where clustering algorithms group spikes into units. Templates are then computed for each unit to match events throughout the recording \cite{buccino2022spike}. Template-matching improves performance in dense recordings and enhances the resolution of overlapping spikes. A typical pipeline is shown in Figure \ref{fig:spikeSortingPipeline}.

Spike sorting has progressed from manual clustering methods to increasingly automated algorithms that can handle dense, long-duration recordings with minimal human intervention. A key breakthrough came with the implementation of automated merging of oversplit units and drift correction, which improved the ability to track units over time in challenging recording conditions \cite{steinmetz2021neuropixels, pachitariu2024kilosort3}. Several groups have expanded on these advances using different strategies. For example, some have modeled continuous probe motion to maintain unit identity across hours-long recordings \cite{boussard2023dartsort}, while others developed parameter-free clustering techniques to partition large-scale arrays for high-throughput spike sorting \cite{diggelmann2018automatic}. Deep learning has also been applied to improve spike detection and resolve overlapping waveforms in densely packed neural recordings \cite{lee2017yass}. In parallel, researchers have introduced graph-based clustering approaches that do not rely on static waveform templates, enabling unsupervised sorting and automatic curation of units during post-processing \cite{pachitariu2024spike}.

Recently, spike sorting has moved toward real-time algorithms that can classify units during ongoing recordings, enabling immediate feedback and integration into closed-loop experiments \cite{van2024rt}. This shift is especially important for SBI applications, where neural activity must be interpreted and responded to in real-time during adaptive learning tasks or neuro-computational control.

After sorting, curation is the next important step. Metrics such as firing rate, signal-to-noise ratio, and refractory violations are computed post hoc to assess unit quality \cite{jain2025unitrefine}. Interactive tools such as Phy, CellExplorer, and Deephys assist researchers in manually validating and refining results \cite{petersen2021cellexplorer, hornauer2024deephys}. Platforms like SpikeForest support cross-sorter consensus and benchmarking \cite{magland2020spikeforest}. Though properly annotated ground truth remains scarce, optotagging \cite{lakunina2020somatostatin}, expert-labeled datasets \cite{jain2025unitrefine}, and synthetic recordings \cite{buccino2021mearec} help validate spike sorting pipelines. Future work may combine quality metrics with task-specific representations, as seen in electrophysiology-based cell type classification \cite{gonzalez2025hippie, beau2025deep}.

\subsection{Standardized Data Formats and FAIR Principles}

With the shift toward large-scale, multimodal neuroscience, standardized data formats and metadata have become essential to ensure reproducibility, interoperability, and integration across platforms. Neurodata Without Borders (NWB) and the Brain Imaging Data Structure (BIDS) have emerged as two leading standards for organizing electrophysiology and neuroimaging data, respectively \cite{rubel2022neurodata, gorgolewski2016brain}. These formats support rich metadata structures, modular extensibility, and compatibility with high-throughput pipelines. Both frameworks are being actively extended to accommodate additional modalities and experimental paradigms. For example, BIDS has evolved from its original neuroimaging focus to include other data types \cite{gorgolewski2016brain}, while NWB provides a flexible HDF5-based architecture tailored to the complex demands of extracellular electrophysiology \cite{rubel2022neurodata}. Together, they support the implementation of the FAIR principles, which call for data to be Findable, Accessible, Interoperable, and Reusable \cite{wilkinson2016fair}.
These standardized formats are now being widely adopted by large-scale neuroscience consortia and data-sharing platforms, facilitating collaborative analysis, benchmarking, and the development of generalizable machine learning tools. As SBI and NeuroAI increasingly rely on multimodal, cross-laboratory datasets, such standards will be essential to integrate experimental data with computational models, including digital twins and neurosymbolic frameworks.

For SBI and OI, where data include electrophysiology, imaging, and behavioral interactions, standardized ontologies and metadata are critical. Variables such as organoid age, cell line, training history, and electrode configuration must be documented. Community-driven projects are working to expand schemas to capture this complexity \cite{smirnova2023organoid}.

Data repositories such as the Distributed Archives for Neurophysiology Data Integration (DANDI) support cloud-based access and analysis of NWB/BIDS data, facilitating collaboration and scalability \cite{subash2023comparison}. Efforts such as the NIH BRAIN Initiative, the International Brain Laboratory, and the Baltimore Declaration for OI emphasize the importance of standardized, sharable data \cite{insel2013nih, abbott2017international, hartung2023baltimore}. Together, these efforts are transforming electrophysiological data from isolated lab results into a cumulative resource for global neuroscience and computational modeling.

\section{Ethical and Societal Considerations}
\label{sec:ethical_issues}
The development of brain organoids has raised important ethical concerns regarding their potential moral relevance \cite{chen2019transplantation, greely2021human, hoppe2023human}. These concerns have intensified as organoids demonstrate spontaneous neural activity that resembles some patterns observed during early human brain development \cite{hernandez2025self, sharf2022functional, van2025protosequences, trujillo2019complex, el2024human}. A central question is whether such constructs could support conscious experience or other morally relevant states \cite{kagan2024embodied}. For example, Integrated Information Theory (IIT) suggests that consciousness may emerge from systems that exhibit complex, integrated patterns of information processing \cite{tononi2012integrated}. Under this framework, organoids capable of memory formation or environmental responsiveness might be considered to possess primitive forms of consciousness that require moral considerations. While frameworks like IIT are far from universally accepted, calls for ethical considerations have grown more pressing as researchers implement closed-loop systems that provide organoids with sensory feedback during learning experiments \cite{kagan2022vitro, robbins2024goal, rountree2025long, lavazza2023human, kataoka2024moral}. These advances intersect with existing challenges related to donor consent and privacy for how donated tissue can and should be used \cite{kataoka2025beyond}. For instance, if human organoids are derived from stem cell lines that originate from donors who were not informed about the potential use of their cells in organoid intelligence research, this may be a concern \cite{hartung2024brain}. As the scope of organoid applications expands, it is essential that donors are informed about the possibility that their cells may contribute to systems designed to process information or acquire learning capabilities. Furthermore, safeguards established for broader research using donated tissue must be strictly enforced to protect privacy, especially if organoids preserve genetically encoded features that could reveal information about a donor’s cognitive or neurological traits \cite{boers2018consent}.

Beyond the ethical considerations that apply broadly to the use of any tissue engineering and/or stem cell-based technologies, some specific issues relating to SBI and OI exist. Therefore, proponents of SBI and OI are actively engaged in managing the ethical implications of developing these new technologies (\cite{smirnova_organoid_2023, kagan2023technology, kagan2022vitro}. The initial formalization of OI proposed an embedded ethics approach to allow for ongoing research \cite{smirnova_organoid_2023, morales2023first}. Actionable ethical frameworks for SBI have also been developed as part of an anticipatory governance process \cite{kagan2023technology}. In particular, three key steps were proposed in this framework. First, clear nomenclature guidelines to describe the technologies and system properties must be resolved. This work has begun and will require a multidisciplinary community-based approach \cite{kagan2024toward}. Second, reliable objective metrics that mark out morally relevant traits should be developed and tested. One approach has been termed 'experimental neuroethics' and offers a pathway for iteratively exploring measurable properties of cultured neural systems across various scales \cite{kagan2024embodied}. Further work related to this has proposed four moral-status-conferring criteria: evaluative stance, self-directedness, agency, and other-directedness, which could apply to organoids if they exhibit evidence of subjective experience \cite{boyd2024moral,boyd2024dimensions}. Finally, the ethical obligations if \textit{in vitro} neural systems do display morally relevant traits need to be determined. Considerations such as the subsidiary principle are relevant, as even if some degree of morally relevant traits arose in some more advanced OI system than is currently available (as no evidence supports these in existing applications), it may still be that the ethical good of avoiding animal use and finding treatments for diseases outweighs the ethical risks \cite{kataoka2025beyond}. Such discussions are well captured in the anticipatory governance approach and will occur more effectively should the need arise \cite{kagan2023technology}.      

Ultimately, these ethical considerations are not solved issues, but requirements for proactive engagement in the evolving technological landscape. Key steps such as proactive monitoring protocols to detect signatures of sentience, the refinement of donor consent practice, and the development of ethical guidelines that scale with system complexity \cite{kagan2023technology,kagan2024embodied} require major efforts and multidisciplinary coordination \cite{lavazza2023human, hartung2024brain, kataoka2025beyond}. As the boundary between biological entities and computational tools becomes less distinct, interdisciplinary collaboration among neuroscientists, ethicists, legal experts, policy makers, and public stakeholders will be essential to ensure responsible progress in this field \cite{mostajo2022emergence, greely2024implementing}.

\section{Recent Advances and Applications in SBI}\label{sec:advances}

Recent advances in SBI, with OI as a key subset, include the development of biohybrid platforms that combine living neural tissue with computational systems for programmable information processing. Excellent groups, projects and programs include but are not limited to the following:

\emph{Cortical Labs} demonstrated that networks of approximately 800,000 cultured neurons can learn a Pong-like task within minutes by minimizing prediction error under the Free Energy Principle \cite{goldwag2023dishbrain, kagan2022vitro, kagan2025harnessing}. This result inspired the CL1 platform, which integrates \textit{in vitro} neural cultures with silicon MEAs in a closed-loop biological intelligence operating system. CL1 supports code deployment for drug discovery, disease modeling, and task-based learning with orders-of-magnitude lower power consumption \cite{kagan2025cl1, watmuff2025drug}.

\emph{Finalspark's Neuroplatform} provides a secure bio-silicon hybrid environment that offers continuous online access to neural organoids. Researchers can remotely interact with the system to conduct experiments in a contained and stable setup \cite{jordan2024open, singh2024development}.

At the \emph{University of California Santa Cruz}, the Braingeneers group is developing neural organoid platforms that interface with both electrical and microfluidic systems \cite{seiler2022modular, voitiuk2024feedback, elliott2023internet, elliott2024pathological}. These tools aim to dissect the role of electrochemical signaling in learning and behavior and to build predictive models of network function \cite{sharf2022functional, andrews2024multimodal, elliott2024pathological, geng2024multiscale, hernandez2025self}. The platform also serves as a foundation for the \emph{Mathematics of the Mind} course, the first academic offering in which undergraduate students directly interact with brain organoids \cite{elliott2023internet}. This educational initiative has recently expanded to include high school students \cite{vera2025reducing}, providing unprecedented access to living neural systems in a learning environment.

\emph{Indiana University} researchers introduced Brainoware, a reservoir computing platform based on organoid-MEA constructs. This system harnesses intrinsic neuroplasticity to dynamically reshape network dynamics and perform temporal pattern recognition with minimal training data \cite{cai2023brain}.

\emph{The University of Illinois Urbana-Champaign’s Mind In Vitro Initiative} offers an open-source electrophysiology suite with modular MEA modules, reconfigurable electrical, optical, and fluidic signal chains, and software for data acquisition and cloud integration. This platform reduces cost by over tenfold and enables scalable access to SBI experimentation \cite{zhang2024mind}.

In parallel, national and international programs are supporting large-scale SBI development. In the United States, the NSF’s Emerging Frontiers in Research and Innovation (EFRI) initiative launched the \emph{Biocomputing through EnGINeering Organoid Intelligence (BEGIN OI)} program. This effort supports multidisciplinary teams in integrating human iPSC-derived organoids with high-density MEAs and adaptive FPGA controllers for real-time learning, low-power computation, and therapeutic screening \cite{nsf_efri_biocomputing}. In Europe, the \emph{Neu-CHiP} consortium combines living cortical neurons with spike-based neuromorphic chips. This bidirectional system allows hardware-generated spikes to modulate neuronal firing and neural outputs to update chip weights, demonstrating closed-loop control and high-fidelity data processing \cite{neuchip}.

Together, these efforts reflect the growing maturity of SBI and OI platforms. By combining biological adaptability with engineered systems, these technologies offer a path toward intelligent systems that compute through real-time interactions between neurons, hardware, and software.

\subsection{SBI Applications in Disease Modeling and Therapeutics}

Brain organoids are central to emerging SBI systems due to their ability to model disease-relevant circuit phenotypes in a human-specific context. In neurodevelopmental disorders such as Autism Spectrum Disorder and schizophrenia, organoids derived from patient iPSCs or CRISPR-edited lines exhibit altered progenitor dynamics, excitatory/inhibitory imbalance, and circuit-level desynchronization \cite{paulsen2022autism, bedford2025brain, neo2023resting, zhang2023human, sebastian2023schizophrenia, villanueva2023advances, howes2024schizophrenia}. These features compromise signal integration and learning capacity, offering testbeds to evaluate closed-loop feedback, plasticity, and pharmacological modulation in SBI platforms.

In epileptic disorders such as tuberous sclerosis complex, Dravet syndrome, and CDKL5 deficiency, organoids exhibit hyperexcitability, disrupted synchrony, and seizure-like dynamics \cite{blair2018genetically, mzezewa2025abnormalities, eichmuller2022amplification, blair2020new, doorn2023silico, glass2024excitatory, wu2022neuronal}. These models expose the boundaries of stability and adaptability in BNN-based computing systems. Similar challenges appear in Rett syndrome organoids with \textit{MECP2} mutations, where reduced network complexity can be used to test closed-loop rescue strategies \cite{nieto2020human, trujillo2021pharmacological, osaki2024early}.

Neurodegenerative models are also related to SBI. Alzheimer’s disease organoids show amyloid deposition and tau pathology \cite{bubnys2022harnessing, fyfe2021brain, papaspyropoulos2020modeling, zhao2020apoe4}, while Parkinson’s disease and Huntington’s disease models capture dopaminergic and GABAergic circuit loss \cite{kim2025parkinson, smits2020midbrain, zheng2023human, zheng2025efficacy, wu2024construction, chen2022human}. These features can allow evaluation of memory retention, pattern recognition, and adaptive behavior in compromised networks.

Multimodal tools such as Patch-seq \cite{cadwell2016electrophysiological}, optotagging \cite{lakunina2020somatostatin}, and electro-seq \cite{li2023multimodal} combined with analytical frameworks like HIPPIE \cite{gonzalez2025hippie} and DeepNeuron \cite{beau2025deep} enable cell-type-specific characterization of functional deficits. These readouts can be integrated into SBI systems to identify, monitor, and compensate for disease-specific vulnerabilities.

While various challenges remain \cite{zhao2024brain, andrews2022challenges}, innovations in vascularization, co-culture, and organoid assembly are improving physiological fidelity \cite{tan2023vascularized, pham2018generation, mansour2018vivo, park2023ips, gerasimova2025novel, kanton2022human, dong2021human, kelley2024host}. Together, disease-specific organoids provide not only mechanistic insight but also dynamic substrates for next-generation biohybrid platforms.

\section{Conclusion}
NeuroAI represents a converging frontier of neuroscience, AI, and biomedical engineering. This review has outlined a conceptual framework that organizes the field into three interacting domains: wetware, which includes living neural substrates such as organoids and cultured networks; hardware, which encompasses neuromorphic systems and BCI interfaces; and software, which spans learning models from deep neural networks to neuro-symbolic AI, RL, and active inference. While each of these domains contributes valuable capabilities, the most transformative potential lies in their integration \cite{kagan2025harnessing}.

SBI, as a representative of this integration, offers a path toward adaptive, energy-efficient, and interpretable intelligence. Realizing this vision will require advances in real-time feedback control \cite{voitiuk2024feedback, voitiuk2021light}, scalable organoid interfacing \cite{kagan2025cl1, park2024modulation, parks2022iot}, standardized multimodal datasets \cite{hernandez2025self, andrews2024multimodal}, and new methods for decoding and guiding emergent dynamics \cite{geng2024multiscale, magland2025facilitating}. Software systems must evolve to interact meaningfully with biological substrates, not just through classification or prediction, but by shaping and co-adapting with living circuits.

As these systems grow in complexity and function, ethical and regulatory frameworks must also evolve to address questions of sentience, consent, and biological data security. With coordinated effort across disciplines, NeuroAI has the potential to redefine how we understand intelligence and build unprecedented intelligent systems. By integrating multi-omics data, leveraging advanced sensing, computing, and interfacing technologies, and embracing a multi-disciplinary approach, researchers can unlock the full potential of these bio-computational systems for basic research and modern medicine \cite{tanveer2025starting}.

\clearpage
\section*{Acknowledgments}
J.G.-F. and M.A.M.-R. were supported by the following grants: Schmidt Futures (SF857);  National Human Genome Research Institute (RM1HG011543); National Science Foundation (2515389); California Institute for Regenerative Medicine (DISC4-16285 and DISC4-16337); University of California Office of the President (M25PR9045); National Institute of Mental Health (U24MH132628 and U24NS146314). The content is solely the responsibility of the authors and does not necessarily represent the official views of the National Institutes of Health, the National Science Foundation,  CIRM, or any other agency of the State of California.

\section*{Competing Interests}
J.G.-F. and M.A.M.-R. are named inventors in patent applications relating to data processing of high throughput electrophysiology. M.A.M.-R. is an advisor for Atoll Financial Group. B.J.K., A.L. are employees of Cortical Labs, a research-focused startup working in a related space and holding patents related to this manuscript. B.J.K. holds shares or another interest in Cortical Labs.

\section*{Author Contributions}
D.P. and M.S.T. contributed equally to this work. D.P. and M.S.T. were responsible for conceptualization, outlining, and drafting the manuscript. J.G.-F. and A.L. contributed to writing specific sections and assisted in literature review and synthesis. B.J.K. and M.A.M.-R. provided critical revisions, factual verification, and substantial editorial input, including improving the structure and clarity of the review. G.W. supervised the project and performed the final revisions and polishing of the manuscript.

\clearpage
\section*{Supplemental Tables}
\setcounter{table}{0}
\renewcommand{\thetable}{S\arabic{table}}

\subsection*{S1. Field Relations in the NeuroAI Hierarchical Graph}

\begin{small}

\begin{longtable}{|p{4cm}|p{8cm}|}

\hline
\textbf{Field Name} & \textbf{Brief Description} \\ 
\hline
\endfirsthead

\hline
\textbf{Field Name} & \textbf{Field Description} \\ 
\hline
\endhead

Neuroscience (NS) & The study of the nervous system, including brain function, neural circuits, plasticity, and cognitive processes \cite{squire2012fundamental}. \\ \hline
Cell \& Tissue Engineering (CTE) & Engineering and cultivating biological tissues and 3D organoids for research, therapeutic applications, and experimental neuroscience \cite{van2022tissue, hofer2021engineering}. \\ \hline
Bioelectronics (BE) & The integration of electronics with biological systems to develop neural interfaces, biosensors, and stimulation devices for bidirectional communication with neural tissue \cite{willner2006bioelectronics}. \\ \hline
Artificial Intelligence (AI) & The development of algorithms and computational systems that simulate aspects of human intelligence, including learning, reasoning, and problem-solving. \\ \hline
Data Science (DS) & The extraction, analysis, and interpretation of large-scale data, including biomedical and neuroscience datasets, to uncover patterns and derive insights \cite{russell2016artificial}. \\ \hline
Computational Biology (CB) & The application of computational methods to analyze, model, and simulate biological processes, systems, and data \cite{waterman2018introduction}. \\ \hline
Synthetic Biology (SB) & The synthesizing of new biological systems or the modification of existing ones to enable functions and properties not present in the system naturally \cite{serrano2007synthetic}. \\ \hline
Tissue Culture Computational Modeling (TCCM = CTE + CB) & The integration of tissue engineering and computational biology to simulate, analyze, and design biological tissue structure, including neural tissues \cite{montes2019mathematical}. \\ \hline
Neural Tissue Culturing (NTC = NS + CTE + SB) & Combining neuroscience, tissue engineering and synthetic biology to cultivate neural cultures with structured cyto-architecture and functional neural activity \cite{lokai2023review}. \\ \hline
Cognitive and Systems Neuroscience (CSN = NS + CB + AI) & The study of brain functions and neural systems using computational modeling, AI, and systems neuroscience to understand cognition, learning, and adaptation \cite{gazzaniga2009cognitive,metzler2012systems}. \\ \hline
Neuroinformatics (NI = NS + CB + DS) & The intersection of neuroscience, computational biology, and data science to manage, analyze, and model complex neural and electrophysiological data \cite{arbib2001computing}. \\ \hline
Machine Learning (ML = AI + DS) & The intersection of AI and data science to develop algorithms that learn from data and improve performance without explicit programming \cite{bishop2006pattern}. \\ \hline
Neuromorphic Architecture (NMA = BE + NS) & The design of hardware that mimics the structure and function of biological neural networks, utilizing event-driven processing and in-memory computing \cite{abdallah2022neuromorphic}. \\ \hline
Neural Tissue Modeling (NTM = NI + CSN + TCCM) & The use of neuroinformatics, cognitive systems neuroscience, and computational modeling to simulate neural cultures' structural and functional dynamics \cite{montes2019mathematical,poli2019experimental}. \\ \hline
Neural Learning Models (NLM = CSN + ML) & AI models inspired by neuroscience, integrating symbolic reasoning and neural learning mechanisms to improve adaptability, interpretability, and cognitive-like processing. Also includes task-oriented models such as reinforcement learning and active inference \cite{garcez2023neurosymbolic,wiering2012reinforcement,friston2016active}.\\ \hline
Neural Tissue Interfacing (NTI = NTC + BE) & The combination of neural culture engineering and bioelectronics to develop interfaces for recording, stimulating, and communicating with neural tissues. \cite{passaro2021electrophysiological}. \\ \hline
Neuromorphic AI (NMAI = NMA + ML) & AI algorithms inspired by the structure and function of biological brains, optimized for neuromorphic computing architectures to achieve efficient, event-driven learning \cite{ivanov2022neuromorphic,schuman2022opportunities}. \\ \hline
Feedback-Driven Learning (FDL = NTI + NLM) & A framework integrating neural interfaces, stimulation, and cognitive neuroscience to enable adaptive, real-time feedback learning in neural tissues. \cite{kagan2022vitro, robbins2024goal}. \\ \hline
Reservoir Computing (RC = NTI + NI + ML) & The application of BNNs as computational reservoirs, leveraging their dynamic neural activity for real-time information processing and learning \cite{cai2023brain, smirnova2023reservoir}. \\ \hline
Neuro Digital Twin Modeling (NDTM = NTM + NSA + NMAI) & The development of a high-fidelity digital replica (digital twin) of neural tissue cultures and organoids, integrating neuro-symbolic AI and computational modeling for simulation and analysis \cite{moller2021digital}. \\ \hline
Synthetic Biological Intelligence (SBI = FDL + RC) & The utilization of BNNs as adaptive learning systems, leveraging real-time feedback and reservoir computing to perform cognitive-like tasks \cite{smirnova2023organoid}. \\ \hline
NeuroAI (NeuroAI = NDTM + SBI) & The ultimate integration of digital twin modeling and SBI to create bio-silico hybrid systems capable of learning, adaptation, and cognitive processing \cite{panuccio2016intelligent, smirnova2024promise}. \\ \hline
\label{tab:oi_fields}
\end{longtable}
\end{small}

\clearpage
\begin{landscape}
\subsection*{S2. Final comparison of major neuromorphic architectures*}
\label{tab:neuromorphic_comparison}
\noindent\begin{minipage}{\linewidth}
  \centering
  \footnotesize
  \begin{tabularx}{1\linewidth}{@{} p{1.25cm} *{6}{Y} @{}}
    \toprule
    \textbf{Aspect} & \textbf{Loihi \cite{davies2018loihi}} & \textbf{TrueNorth \cite{akopyan2015truenorth}} & \textbf{SpiNNaker \cite{furber2012overview}} & \textbf{DYNAP-SE2 \cite{richter2024dynap}} & \textbf{BrainScaleS2 \cite{pehle2022brainscales}} & \textbf{Tianjic \cite{deng2020tianjic}} \\
    \midrule
    Clocking
      & Fully async; event-driven (no global clock)
      & Mixed async–sync cores; event-driven routing
      & Asynchronous interconnect; no global clock
      & Fully async mixed-signal design
      & 1{,}000$\times$ accelerated analog time evolution; digital processors for control
      & Hybrid ANN/SNN; sync cores + async mesh routing
    \\
    Communica-tion
      & 2D-mesh NoC; AER-based spike routing
      & 2D-mesh; asynchronous router; AER routing; merge–split blocks for multi-chip scaling
      & Packet-based NN, P2P, MC, FR packets; 2D torus NoC
      & On-chip AER; 1024 neurons per chip
      & On-chip event-routing network + custom off-chip link protocol
      & 2D-mesh NoC for spikes and ANN data
    \\
    Neuron model
      & Digital LIF; multi-compartment; stochastic options
      & Augmented LIF; stochastic + deterministic, 1–3 neurons emulate 20 spiking behaviors
      & Point neuron model; 1000 neurons/core real-time
      & Thresholded I\&F and AdEx; subthreshold analog
      & Analog adaptive Exponential IF (AdEx)
      & Hybrid: binary SNN neurons; 8-bit ANN activations
    \\
    Synapse type
      & On-chip SRAM; 1–9 bit weights; 16\,MB total
      & 64k synaptic crossbar (256×256), four weight types (integer)
      & Excitatory \& inhibitory with scalar weights
      & 64 synapses; 4-bit weight; 2-bit delay; STP depression
      & SRAM-stored weights; current \& conductance-based synapses
      & SRAM per core; shared ANN/SNN weights
    \\
    On-chip learning
      & Yes: microcoded STDP \& RL plasticity
      & No: inference only; offline training
      & Yes: in software on ARM cores
      & Local short-term plasticity + homeostasis; no long-term plasticity
      & Yes: PPUs for STDP, R-STDP, homeostatic, structural plasticity
      & No: inference only; pre-trained weights
    \\
    Energy per syn-op
      & 20–25\,pJ; STDP $\sim$120\,pJ
      & 400 GSOPS/W (max); 46 GSOPS/W at 65 mW for typical workload
      & 2{,}200 MIPS/W (1{,}036{,}800 cores $\times$ 200\,MHz / 90\,kW power)
      & 150 pJ (I\&F); 300 pJ (AdEx) at 80 Hz
      & N/A
      & 1.28 TOPS/W (ANN); 649 GSyOPS/W (SNN) at 300 MHz, 0.85 V
    \\
    Biological realism
      & Spiking, local plasticity, moderate realism
      & Spiking dynamics approximate biological behavior; 1–3 neurons replicate 20 Izhikevich behaviors
      & Real-time; simplified neuron/synapse models
      & Modeled Alpha-EPSCs, NMDA gating, AMPA diffusion, adaptation
      & High: multi-compartment analog neurons + biologically plausible learning
      & Hybrid SNN/ANN; spiking dynamics + ANN activations
    \\
    \bottomrule
    \addlinespace[1pt]
    \multicolumn{7}{>{\raggedright\arraybackslash}p{\dimexpr\linewidth-2\tabcolsep\relax}}{%
      \tiny \textbf{*}NoC = Network-on-Chip; AER = Address-Event Representation; P2P = Point-to-Point; MC = Multicast; FR = Fixed-Route; LIF = Leaky Integrate-and-Fire; I\&F = Integrate-and-Fire; AdEx = Adaptive Exponential Integrate-and-Fire; R-STDP = Reward-modulated STDP; RL = Reinforcement Learning; PPU = Plasticity Processing Unit; EPSC = Excitatory Post-Synaptic Current; AMPA = $\alpha$-amino-3-hydroxy-5-methyl-4-isoxazolepropionic acid; TOPS/W = Tera-Operations per Second per Watt; GSOPS/W = Giga Synaptic Operations per Second per Watt; GSyOPS/W = Giga \emph{Spiking} Synaptic Operations per Second per Watt; MIPS/W = Million Instructions per Second per Watt; SRAM = Static RAM; ARM = Advanced RISC Machine.%
    }\\
  \end{tabularx}
\end{minipage}

\end{landscape}

\newpage
\bibliography{sn-bibliography}
\end{document}